\pdfoutput=1
\documentclass{article}

\usepackage{ifthen}
\usepackage{amssymb}
\usepackage{amsmath}
\usepackage{graphicx}
\usepackage{algorithm}
\usepackage{algpseudocode}
\usepackage{hyperref}
\usepackage{amsthm}

\usepackage[final]{changes}
\definechangesauthor[name=R1, color=orange]{R1}

\newcommand{\TheTitle}{Delayed approximate matrix assembly in multigrid with dynamic precisions}

\newboolean{specialissue}
\newboolean{arxiv}

\setboolean{specialissue}{false}
\setboolean{arxiv}{true}

\ifthenelse{\boolean{specialissue}}{
  \articletype{Article Type}
  
  \received{26 April 2016}
  \revised{6 June 2016}
  \accepted{6 June 2016}

  \raggedbottom
}{}

\ifthenelse{\boolean{arxiv}}{
  \usepackage{geometry}

\title{\TheTitle 
  \thanks{
 		The work was funded by an EPSRC DTA PhD scholarship (award
 		no.~1764342).
  		It made use of the facilities of the Hamilton HPC Service of Durham
  		University.   		
  		The underlying project has received funding from the European Union's Horizon 2020 research and innovation programme under grant agreement No 671698 (ExaHyPE). 
  		This paper is an extended version of C.D.~Murray and T.~Weinzierl:
  		\emph{Lazy stencil integration in multigrid algorithms} as introduced and
  		published at the PPAM'19 conference.
  }
}

\newenvironment{observation}{}{}

\author{
  Charles D.~Murray \and
  Tobias Weinzierl
}

}{}

\begin{document}

\ifthenelse{\boolean{specialissue}}{

\title{
  \TheTitle 
  \protect\thanks{
 		The work was funded by an EPSRC DTA PhD scholarship (award
 		no.~1764342).
  		It made use of the facilities of the Hamilton HPC Service of Durham
  		University.   		
  		The underlying project has received funding from the European Union's Horizon 2020 research and innovation programme under grant agreement No 671698 (ExaHyPE). 
  		This paper is an extended version of C.D.~Murray and T.~Weinzierl:
  		\emph{Lazy stencil integration in multigrid algorithms} as introduced and
  		published at the PPAM'19 conference.
  }
}

\author[1]{Charles D.~Murray}
	
\author[1]{Tobias Weinzierl}

\authormark{C.~D. Murray and T.~Weinzierl}

\address{
 \orgdiv{Department of Computer Science}, 
 \orgname{Durham University}
}

\corres{
  Charles D.~Murray, 
  \email{c.d.murray@durham.ac.uk}
}

\presentaddress{
 Department of Computer Science,  
 Durham University, 
 Lower Montjoy, South Road,
 DH1 3LE Durham, GREAT BRITAIN
}

  \abstract[Summary]{
    The accurate assembly of the system matrix is an important step in any
code that solves partial differential equations on a mesh.
We either explicitly set up a matrix, or we work in a matrix-free environment
where we have to be able to quickly
return matrix entries upon demand.
Either way, the construction can become costly due to
non-trivial material parameters entering the equations,
multigrid codes requiring
cascades of matrices that depend upon each other, or
dynamic adaptive mesh refinement that necessitates the
recomputation of \replaced[id=R1]{matrix entries or the whole
equation system}{matrices} throughout the solve.
We propose that these constructions can be performed
concurrently \replaced[id=R1]{with the multigrid cycles}{with the solve itself}.
Initial geometric matrices and low accuracy integrations 
kickstart the \replaced[id=R1]{multigrid}{solve}, while 
improved assembly data is fed to the solver as and when
it becomes available.
The time to solution is improved as we eliminate an expensive preparation phase
traditionally delaying the actual computation.
We eliminate algorithmic latency.
Furthermore, we desynchronise the assembly \deleted[id=R1]{process} from the
\replaced[id=R1]{solution process}{solve}.
This anarchic increase of the concurrency level improves the scalability.
Assembly routines are notoriously memory- and bandwidth-demanding.
As we work with iteratively improving operator accuracies, 
we finally propose the use of a hierarchical, lossy compression scheme
\deleted[id=R1]{similar to multiprecision storage} such that the memory
footprint is brought down aggressively where the system matrix entries carry little information
or are not yet available with high accuracy.

  }

  \keywords{
   Finite element assembly, dynamically adaptive Cartesian grids,
   algebraic-geometric multigrid, delayed operator computation, mixed precision
   computing, asynchronous multigrid
  }
}{
}

\maketitle

\ifthenelse{\boolean{arxiv}}{
  The accurate assembly of the system matrix is an important step in any
code that solves partial differential equations on a mesh.
We either explicitly set up a matrix, or we work in a matrix-free environment
where we have to be able to quickly
return matrix entries upon demand.
Either way, the construction can become costly due to
non-trivial material parameters entering the equations,
multigrid codes requiring
cascades of matrices that depend upon each other, or
dynamic adaptive mesh refinement that necessitates the
recomputation of \replaced[id=R1]{matrix entries or the whole
equation system}{matrices} throughout the solve.
We propose that these constructions can be performed
concurrently \replaced[id=R1]{with the multigrid cycles}{with the solve itself}.
Initial geometric matrices and low accuracy integrations 
kickstart the \replaced[id=R1]{multigrid}{solve}, while 
improved assembly data is fed to the solver as and when
it becomes available.
The time to solution is improved as we eliminate an expensive preparation phase
traditionally delaying the actual computation.
We eliminate algorithmic latency.
Furthermore, we desynchronise the assembly \deleted[id=R1]{process} from the
\replaced[id=R1]{solution process}{solve}.
This anarchic increase of the concurrency level improves the scalability.
Assembly routines are notoriously memory- and bandwidth-demanding.
As we work with iteratively improving operator accuracies, 
we finally propose the use of a hierarchical, lossy compression scheme
\deleted[id=R1]{similar to multiprecision storage} such that the memory
footprint is brought down aggressively where the system matrix entries carry little information
or are not yet available with high accuracy.

}{}

\section{Introduction}
\label{section:introduction}

%
%
Multigrid algorithms are among the fastest solvers known for
elliptic partial differential equations (PDEs) of the type
%

\begin{equation}
 - \nabla \left( \epsilon \nabla \right) u = f
 \label{introduction:equation}
\end{equation}


\noindent
on a $d$-dimensional, well-shaped domain $\Omega $.
An approximation of the function $u: \Omega \mapsto \mathbb{R}$ is what we are
searching for with
$\epsilon : \Omega \mapsto \mathbb{R}^+$ as a material parameter,
and $f: \Omega \mapsto \mathbb{R}$ constituting the right-hand
side.
The system is closed by appropriate boundary conditions.
We restrict ourselves to Dirichlet conditions here.
A Ritz-Galerkin \replaced[id=R1]{finite element}{Finite Element} discretisation
over a mesh $\Omega _h$ that geometrically discretises $\Omega $ yields an equation system $A_h u_h = f_h$.
This is the \replaced[id=R1]{linear equation system}{equation} actually tackled
by multigrid \added[id=R1]{(MG)}.
Equations of the type (\ref{introduction:equation}) arise in many application
domains studying (quasi-)stationary phenomena.
They also are an important building block within many time-dependent
problems, where they model incompressibility conditions or friction for example.
Finally, they also arise in Lagrangian setups,
where they model gravity between moving objects for example.

Common to all sketched application areas is that solving $A_h u_h = f_h$,
i.e.~finding $u_h \approx A_h^{-1} f_h$, is expensive.
The ellipticity of (\ref{introduction:equation}) implies that any change of a
solution value anywhere within $\Omega $ impacts the solution over
the entirety of $\Omega $.
Finite \replaced[id=R1]{elements}{Elements} and related techniques manage to
break up this strong global dependency by discretising the PDE with test and shape functions that have local support.
Any single update of an entry of $u_h$ affects only neighbouring elements within the
discretisation, i.e.~few other entries within $u_h$.
It propagates through the
whole domain from there.
We work with equation systems that are sparse, and thus manageable from
both a memory and compute effort point of view.
However, information propagation this way is intrinsically slow. 
Multigrid compensates for this effect as it removes errors across a hierarchy of
meshes.
Coarser and coarser grids take ownership of
updates that yield non-local modifications, i.e.~they handle
low-frequency errors from the fine grid.
This multilevel approach to tackle an elliptic problem on a cascade of
resolutions
is the seminal idea behind
multigrid and the reason why multigrid is fast.

\begin{figure}
 \begin{center}
  \includegraphics[width=0.52\textwidth]{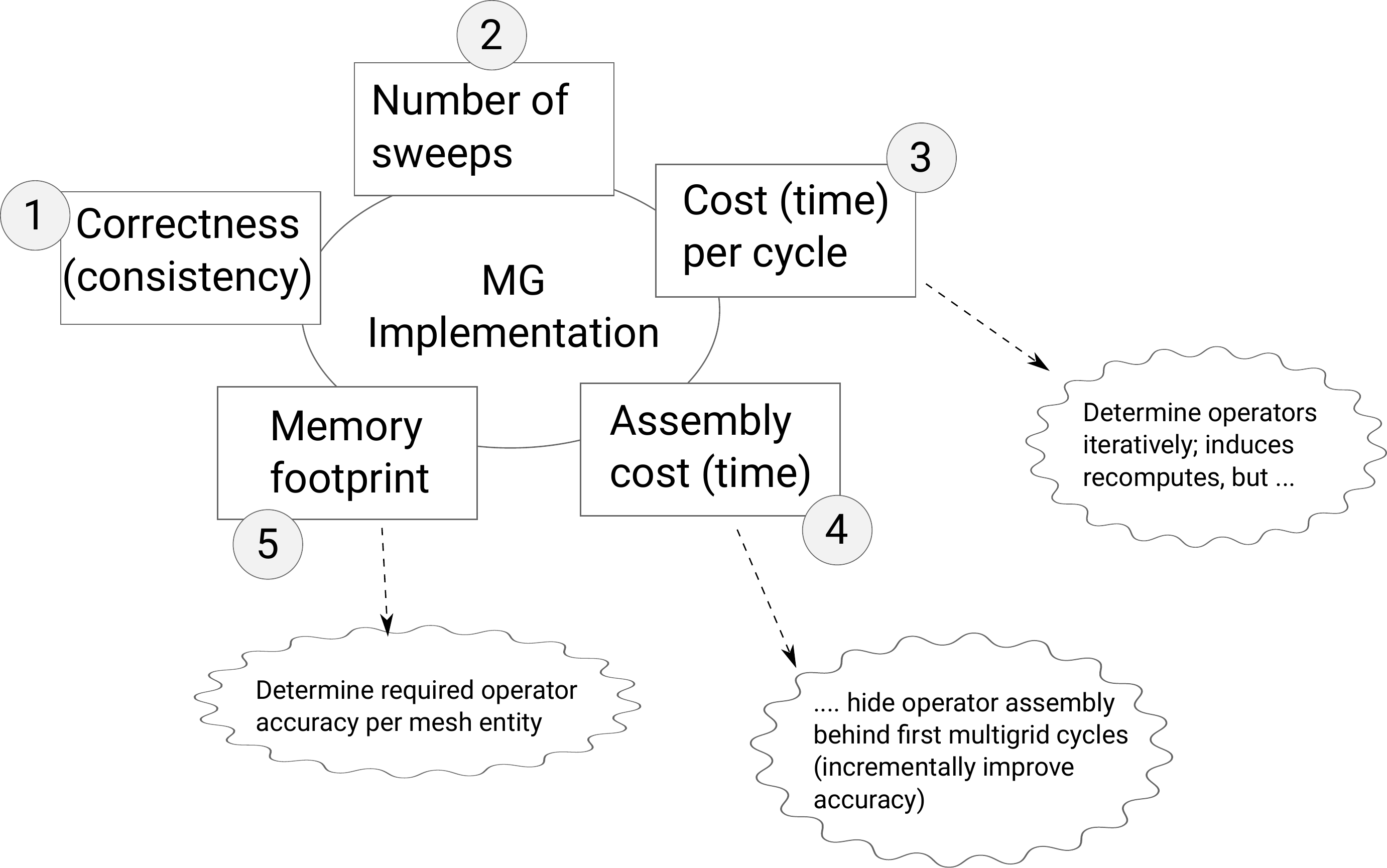}
 \end{center}
 \vspace{-0.4cm}
 \caption{
  \added[id=R1]{
   Multigrid implementation challenges: (1) An implementation has to be
   correct, i.e.~yield the result of the underlying mathematics and thus be
   consistent with it, (2) there have to be as few data structure as possible,
   while the cost per cycle (3), the assembly/setup cost (4) and the memory footprint (5) have to be small, too.
   (3),(4),(5) are the core areas where we make a contribution, while 
   implications on (1) and (2) are studied.
  }
  \label{figure:intro:paper-concept}
 }
\end{figure}

%
%
When we implement multigrid, \added[id=R1]{
 the implementation has to be in line with the mathematical theory, while  
}
the cost of the actual \replaced[id=R1]{implementation}{solve}
is determined by at least four factors:
the number of solver iterations;
the time taken for a single solver iteration \added[id=R1]{including all data
traversals---of which there can be multiple of them in the MG context}; the compute cost (time) to set up all
required operators (matrices) on all of the different scales and all required
operators coupling different scales; \added[id=R1]{and} the memory footprint of
the method \added[id=R1]{(Fig.~\ref{figure:intro:paper-concept})}.
Traditionally, \added[id=R1]{computer science} research focuses on the first
two aspects---both from a numerical point of view and with performance engineering glasses on.
Yet, there are more and more cases where the latter two
constraints become prescient.
There are at least three reasons for this.
Firstly, the proportion of the runtime spent on the
\replaced[id=R1]{solution phase}{actual solve} often diminishes relative to
(re-)assembly time:
If (\ref{introduction:equation}) serves as a building block within a time-stepping
code, a solution $u$ will often not change radically between time steps.
Along the same lines, dynamic adaptivity changes the mesh and consequently
demands for a new discretisation of (\ref{introduction:equation}).
Yet, a grid update usually does not change the solution completely.
As a result, a low number of iterations or cycles yields a valid solution in both cases
as long as we start
from the previous step's solution as an initial guess.
Secondly, we suffer from a widening memory gap on our machines
\cite{Dongarra:14:ApplMathExascaleComputing}.
The increase in memory access speed cannot keep pace with the growth of
compute power on newer systems.
As dynamic adaptivity---and non-linear systems which are out of scope
here---become mainstream, we assemble multiple times\deleted[id=R1]{ per
multigrid solve}.
Conversely one assembly remains sufficient for static meshes and linear PDEs.
Each re-assembly stresses the memory interconnects since it is typically not 
compute-intense.
Yet, this is the step where volumetric domain information
($\epsilon $-fields) has to be streamed into the core.
The impact is amplified by multigrid's inherent multiscale nature: 
It is not only a single matrix $A_h$ discretised (\ref{introduction:equation}), 
rather we have to maintain and construct a series of matrices
$A_h,A_{kh},A_{k^2h},\ldots$ for
\replaced[id=R1]{coarsening by a factor of $k \geq 2$}{$k$-coarsening}, as well
as the corresponding inter-grid transfer operators, i.e.~prolongations and restrictions.
Finally, due to this multitude of involved operators, multigrid is memory-demanding. 
\deleted[id=R1]{
In $d$ dimensions and Cartesian environments, the
actual system matrix stores around $N \cdot 3 ^d$ entries for a simple $3^d$-point stencil and $N$
unknowns.
Adaptivity changes these numbers slightly.
However, multigrid requires up to}
\deleted[id=R1]{
\noindent
entries as we store multiple levels plus the inter-grid transfers. We assume
classic $k$-coarsening and a $d$-linear inter-grid transfer pattern here, while
$\ell =0$ denotes the finest mesh.
Further metadata and temporary quantities such as right-hand sides per level
or stored residual variables make even efficient multigrid solvers introduce a
memory footprint increase of a factor of 2--3 relative to the original
equation.
}
In an era with stagnating or even decreasing memory per code,
\replaced[id=R1]{the}{such an} overhead \added[id=R1]{to store restriction,
prolongation and coarse grid matrices} quickly becomes a limiting constraint.
Purely geometric approaches with rediscretisation avoid this overhead.
\added[id=R1]{
 They construct local matrix entries on-demand; usually when they are required
 throughout a matrix-vector product.
 In such a case, there is no need to store the matrix entries et all.
}
It is however known that massive sudden $\epsilon $ changes, additional convective
terms or nonlinearities here require very detailed operator recomputations.
We trade the memory demands for heavy compute effort.
\added[id=R1]{
 Furthermore, geometric multigrid tends to become unstable.
 in case of non-trivial $\epsilon$ distributions
\cite{Reps:17:Complex,Sampath:08:Dendro,Sampath:10:Parallel} and is not stable
in a multigrid sense either \cite{Weinzierl:18:Quasi}.
}

%
%
The additional operators introduced by multigrid on top of the actual
discretisation are correction
operators, while the overall scheme is an iterative approach.
Furthermore, the fundamental solutions to (\ref{introduction:equation})
quickly decay.
Ergo, neither do 
the coarse grid equations have to tackle the
exact equation right from the start, nor do we need exact fine grid operators
prior to the first iteration.
Furthermore, if an operator is slightly wrong, i.e.~if it yields slightly wrong
iterates, this error will mainly affect the area around the         
erroneous update in subsequent applications, as the fundamental solutions to 
(\ref{introduction:equation}) quickly decay.
Therefore, it is sufficient to kick off with approximate operators, as long
as\deleted[id=R1]{:}
(i) we later increase their accuracy such that we eventually solve the
right equation system; (ii) the smoother
continually pushes the solution into the right direction; (iii) the correction
equations do not impede these improvements; and (iv) we quickly get
the majority of operators within the computational domain right. 
\replaced[id=R1]{Our idea therefore is it}{We propose} to kick off multigrid
with very crude fine grid approximations plus geometric coarse grid and inter-grid transfer
operators which can be quickly precomputed.
\deleted[id=R1]{
These operators are unstable in case of non-trivial $\epsilon$ distributions
\cite{Reps:17:Complex,Sampath:08:Dendro,Sampath:10:Parallel} and not stable in a
multigrid sense either \cite{Weinzierl:18:Quasi}.
}
While we run the multigrid cycle, we \deleted[id=R1]{therefore} successively
improve the approximation quality of the fine grid operators.
These improvements are deployed as tasks which run in the background of the
actual \replaced[id=R1]{solver}{solve's computation}:
While our solver determines initial solution
approximations, we derive the correct operators describing
the true solution.
\deleted[id=R1]{
A task diagram representation of the sequential variant and our proposed
modification can be seen in Fig.~\ref{figure:intro:pipelining-illustration}.
}
The overall integration is precision-guided, i.e.~we do continuously improve
those equation system entries that continue to benefit from improved integration
\deleted[id=R1]{accuracy} as well as improved storage accuracy.
There is neither a global, uniform numerical integration precision nor a
uniform global (IEEE) floating point format. 
The whole mindset unfolds its beauty once we stop storing
multigrid's matrices explicitly.
Instead, we embed the matrix entries into our mesh---a strategy we label as
quasi matrix-free \cite{Weinzierl:18:Quasi}---and store solely differences to
geometric operators.
The information within any low accuracy stencil, \replaced[id=R1]{no matter
whether}{if} sufficient or not yet available with higher accuracy, can then be
encoded with few bytes only.
Overall, our contribution is three-fold: 
\begin{enumerate}
  \item 
We allow for a solver start without an expensive pre-solve assembly phase,
i.e.~we eliminate algorithmic latency;
  \item 
we increase the code's concurrency level
as we decouple the actual assembly process from the solve and make it
\deleted[id=R1]{anarchically} feed into the latter\added[id=R1]{ anarchically,
i.e.~as soon as results become available yet without any synchronisation or
temporal ordering};
  \item 
and we \deleted[id=R1]{finally} bring down the memory
footprint\deleted[id=R1]{ for the solver steps dramatically}.
This notably affects the initial steps which tend to be cheap anyway, as dynamic
grid refinement just starts to unroll the real compute grid.
\end{enumerate}

\noindent
To the best of our knowledge, this is the first algorithmic blueprint
systematically exploring how to incorporate problem-dependent numerical integration of fine
grid stencils, adaptive coarse grid operator computation and non-IEEE storage
formats without the cost of any arduous assembly into one multigrid implementation.

%
%
The remainder of the paper is organised as follows:
We first \added[id=R1]{
 discuss potential caveats of our philosophy in Section
 \ref{section:shortcomings} that we want to keep in mind and address throughout
 the paper.
 We then
}
review published and related work, which contextualises the present
work, explains where ideas come from and how the proposed techniques fit to
other activities.
All details required to understand the novel ideas of our work are introduced
together with our multigrid algorithm of interest in Section~\ref{section:mg}.
From there, we establish our notion of a delayed, iterative
stencil assembly \added[id=R1]{
 with flexible precision
} (Section~\ref{section:delayed-iterative}).
We dedicate Section~\ref{section:results} to a discussion of the
solver's properties that result from this novel assembly paradigm.
A brief summary and an outlook close the discussion.

\section{Shortcomings of the proposed concepts}
\label{section:shortcomings}

%
%
There are certain \added[id=R1]{obvious} caveats with our proposed solution:
\begin{enumerate}
  \item \replaced[id=R1]{The}{First, the}
matrix entry computation could be subject to starvation.
If the results of new integration tasks are not dropping in on time, the
multigrid solver might converge towards the wrong solution and prematurely
\replaced[id=R1]{signal terminate}{``think'' that it is already there}.
If the integration is run with high priority and thus precedes
the actual \replaced[id=R1]{solution process}{solve}, it is not subject to these
concerns.
Yet, it \replaced[id=R1]{loses}{looses} the proposed selling point.
One might argue that complex integration patterns typically arise only around
submanifolds within the domain.
Despite our remarks on a rapid decay of errors, the elliptic nature of the
problem of choice however implies that inaccurate integration within a subdomain
can pollute the entire solution.
The starvation phenomenon thus has to be analysed.
  \added[id=R1]{We have to validate that the total number of cycles/iterations
  required is not increased massively (item (2);
  Fig.~\ref{figure:intro:paper-concept}).}
  \item \replaced[id=R1]{The}{Second, the}
motivation behind multigrid's construction---we work with a
cascade of coarser and coarser grids where the coarse grids ``take care'' of
global propagation---implies that any fine
grid stencil modification has its impact felt throughout all grid levels.
As long we change the discretisation, the exact nature of all coarse grid operators
also continues to change.
To avoid repeated data accesses per sweep, we limit the multilevel propagation
speed, i.e.~our operator updates propagate from fine to coarse one
level \replaced[id=R1]{per cycle}{at a time}.
Updates ripple through the system. 
For multiplicative multigrid, such a one-level-at-a-time policy is reasonable,
as the \replaced[id=R1]{solver}{solve} also handles one solution at a time.
However, additive multigrid processes all resolutions in one rush.
Coarse operators thus are incorrect if
operator changes are not immediately rolled out to all
levels.
Therefore we exclusively focus on additive multigrid here.
It is more challenging.
The rippling then has to be anticipated carefully---in particular once we
work with dynamically adaptive meshes---as we run into risks
of coarse grid updates pushing the solution in the wrong
direction\added[id=R1]{, i.e.~would yield inconsistent update rules (item (1);
Fig.~\ref{figure:intro:paper-concept})}.
Non-linear setups would yield the same issues.
  \item \replaced[id=R1]{Our}{Finally, our} techniques are not lightweight in a
  sense that they can easily be added to existing solvers in a black-box fashion.
This particularly holds for the usage of multiple precision
formats\deleted[id=R1]{ within one solver}, for which most software might be
ill-prepared.
  \added[id=R1]{While our experiments focus on structured adaptive meshes
  resulting from spacetrees only, it is clear that they apply directly to
  unstructured meshes even if they lack a built-in hierarchy, since both
  geometric and algebraic coarsening construct a hierarchy anyway. Our 
  accuracy dependencies and recomputation needs follow this hierarchy. It is
  however clear that the integration of the ideas into existing unstructured
  mesh software requires additional effort and software design.
 }
\end{enumerate}

\added[id=R1]{
 \noindent
 Our experiments suggest that the first item is not observed in practice,
 even though our work studies a worst-case setup with our additive MG solvers.
 We propose a solution to the second caveat.
 For the third item, our work provides}
\deleted[id=R1]{Yet, they provide} clues for what solver
development roadmaps might \added[id=R1]{want to} incorporate in the future.

\section{Related work and influencing ideas}


Early work by Achi Brandt
\cite{Brandt:79:Multi} 
---specifically his work on MLAT---already clarifies that 
``discretization and solution processes are intermixed with, and greatly benefit from, each other''
in an adaptive scheme.
This principle is not exclusive to MLAT.
It holds for all forms of adaptive mesh refinement\added[id=R1]{ (AMR)}.
If dynamic AMR starts with a coarse discretisation and unfolds the mesh
anticipating the real solution's behaviour, 
we can read this as delaying an exact computation of the fine grid equations
until we know that they are needed.
Dynamic adaptivity also \replaced[id=R1]{mangles}{combines/merges} the assembly
with the \replaced[id=R1]{linear equation system solver}{solve}.

Within the solver world,
the seminal additive scheme proposed by Bramble, Pasciak and Xu \cite{Bramble:90:Parallel,Bastian:98:Additive}
does not utilise exact equation representations on coarse grid levels.
Multiple coarse grid corrections are computed independently from the same fine grid
residual and eventually summed up.
However, the coarse grid correction equation in a multigrid sense results from a
simple $h-$scaling of the fine grid equation's diagonal.
This approximation is sufficient for convergence in many cases \cite{Smith:04:Domain}.
Starting with inexact equations or setups and improving them subsequently is not
new within the multigrid-as-a-solver community either.
Adaptive AMG \cite{Brezina:05:Adaptive,Brezina:06:Adaptive}
constructs coarse grids iteratively.
A tentative coarsening setup is improved by applying it to a set of
candidate vectors for a small number of iterations.
Similarly, \replaced[id=R1]{Bootstrap}{Boostrap} AMG 
\cite{Brandt:11:Bootstrap}
modifies both prolongations and the coarse grid hierarchy itself
by applying them to randomly constructed problems
and modifying them to improve convergence.
In both Adaptive AMG and \replaced[id=R1]{Bootstrap}{Boostrap} AMG,
all modifications of the coarse grid
equations happen during an extended setup phase
rather than during the \replaced[id=R1]{iterations of the solution
process}{actual solve}.

Inaccurate operator approximations are popular when solving non-linear equations.
\replaced[id=R1]{In the}{The} 
Inexact Newton Method
\cite{Dembo:82:Inexact,Martinez:95:Inexact}
for example \replaced[id=R1]{the}{does not
exactly solve the Newton equations in each \replaced[id=R1]{step}{stage of the
non-linear solve}. A} 
Jacobian is approximated once and then used within an iterative process
yielding a sequence of corrections to the solution.
Along the same lines, multigrid for non-linear problems often uses multigrid
within a Newton solve where the non-linear operator is linearised,
i.e.~approximated.
Finally, some multigrid techniques for convection-dominated scenarios symmetrise
the underlying fine grid phenomenon \cite{Yavneh:12:Nonsymmetric} before they
make it subject to algebraic coarsening.
The motivation here is to stabilise the solver, but \added[id=R1]{it} obviously
constructs a regime with inaccurate coarse grid operators.
Our solver overview is far from comprehensive.

For time-dependent problems, many established (commercial) codes  still employ
direct solvers that store an explicit inverse of the system matrix.
This is particularly attractive in scenarios where the mesh does not change, as
the inversion of a matrix preceding a new time step's solution is performed only
once. After that, we merely apply the known inverse, i.e.~rely on a
matrix-vector product.
The massive memory footprint required to store an inverse of a sparse
system makes this approach quickly prohibitive.
However, applying multigrid to each and every time step also 
yields excessive costs, as the solution changes smoothly in time.
We therefore have previously studied a ``multigrid'' concept where a single
level smoothing step is followed by a two-level scheme, followed by another single
level solve, a four level scheme, and so forth \cite{Weinzierl:12:Geometric}.
The coarser a grid level, the more it affects future solutions even though we do
not update the underlying operator in time anymore. 
Such a scheme translates multigrid's $h$-coarsening idea into the time
domain.
Most approaches \replaced[id=R1]{from}{form} the parallel-in-time community (see
\cite{Speck:14:Space} and follow-up work for newer trends) also exploit the idea to approximate the
coarse operators crudely, but to incrementally improve them\deleted[id=R1]{
throughout the solve}.

On the implementation side, a lot of mature software supports matrix-free
solvers today.
PETSc for example phrases its algorithms as if it had a fully assembled matrix
at hand. 
However, many of its core routines make \added[id=R1]{no} assumptions about how
this matrix is actually assembled.
In its MatShell variant, it specifically allows users to deliver matrix parts
on-demand \deleted[id=R1]{through through} \cite{PETSc:Man}, i.e.~PETSc asks
for the matrix part, applies it to the vectors of interest, and immediately discards the operator
again.
While such an approach allows users to (re-)compute all operators whenever they
are required, a sole matrix-free implementation runs the risk to quickly become
inefficient or unstable.
If material parameters in (\ref{introduction:equation}) change rapidly, any
on-demand operator evaluation has to integrate the underlying weak formulation
with high accuracy.
This quickly becomes expensive.
On the coarser grids, a sole matrix-free mindset means that operators cannot
depend on the next finer operator following a Ritz-Galerkin multigrid strategy,
as this next finer operator is not available explicitly either and will
recursively depend on even finer levels.
For these shortcomings, the hybridisation of algebraic and geometric multigrid
is \replaced[id=R1]{well-trodden}{well trod} ground
\cite{Gmeiner:14:HHG,Lu:14:HybridMG,Sundar:12:Parallel,Rudi:15:GordonBell}.
Most hybrid approaches use algebraic multigrid where sole geometric operators
fail, but employ geometric operators wherever possible.
As material parameters change ``infrequently'' on fine meshes and diffusion
dominates in many setups, it is indeed possible to rely on a geometric operator
construction for the majority of the equation systems.

Our own work in \cite{Weinzierl:18:Quasi} introduces an alternative notion of
hybrid algebraic-geometric multigrid.
All operators here are embedded into the
grid and, in principle, algebraic.
As they are encoded within the mesh, we can access them within a matrix-free
mindset.
Storing the operators relative to geometric operators in a compressed form
allows us to work with a memory footprint close to geometric
multigrid.
The present paper follows-up on this strategy and fuses the underlying idea of
lossy compression with iterative operator assembly.
It stores operators with reduced precision where appropriate.
We have previously explored lossy low precision storage both in a multigrid
\cite{Weinzierl:18:Quasi} and an SPH \cite{Eckhardt:15:SPH} context.
Both approaches demonstrate how much we can save in terms of memory footprint,
and both publications point out that the price to pay for this is an increased
arithmetic workload. 
In the SPH context, we propose and prototype how the additional operators can
be deployed to tasks of its own. 
However, the strict causal dependency there implies that performance penalties
can and do arise.
Performance penalties resulting from the increased compute load---it is
notably more expensive to check to which degree we can store an operator
with reduced precision---do not arise in the present case, as we propose an
anarchic scheme where we ignore dependencies of the actual solve on the newly introduced
stencil assembly tasks plus any storage loss analysis.
\deleted[id=R1]{
This is possible as we already work with inaccurate operators.
}


\section{A matrix-free additive multigrid solver on spacetrees}
\label{section:mg}


\replaced[id=R1]{
 Let $ A_h u_h = f_h $ describe
}
{ $ A_h u_h = f_h $ describes }
the equation system that arises from a nodal Ritz-Galerkin 
\replaced[id=R1]{finite element}{Finite
Element} discretisation of (\ref{introduction:equation}).
The test and shape function space are the same.
We use $d$-linear functions.
Hence, we obtain the classic 9-point or 27-point stencils 
on regular meshes.
Each stencil describes the entries of one row of $A_h$.
It describes how a single point (vertex) on one level depends on its cell-connected
neighbours.
The solution vector $u_h$ stores the weights
(scalings) of the individual shape functions.
\added[id=R1]{
 This simple choice of mathematical ingredients allows for a multitude of
 multigrid flavours already.
 We classify some of these flavours and point out which flavours we study in the
 present paper.
 The list is not comprehensive but focuses on solver nuances which are affected
 by the proposed techniques.
}


\added[id=R1]{\emph{Geometric vs.~algebraic construction of multilevel
hierarchy.}} 
Multigrid solvers can be classified into either mesh-based
coarsening or solvers with algebraic coarsening \cite{Brezina:01:Algebraic,Ruge:87:AMG}.
The former rely on an existing cascade of coarser and coarser geometric meshes
$\Omega _h, \Omega _{kh}, \Omega _{k^2h}, \ldots$ which often embed into each
other.
Algebraic coarsening derives the coarse meshes from a connectivity analysis
of $A_h$, i.e.~directly from the matrix without a geometric interpretation.
In our work, we stick to a geometric approach relying on spacetrees
\cite{Weinzierl:Mehl}:
We take the domain $\Omega $ and embed it into a cube.
The cube yields a mesh $\Omega _0$ without any real degrees of freedom, as all
vertices either discretise points outside of the domain or coincide with the
domain boundary.
We cut the cube into $k$ equidistant slices along each coordinate axis to
end up with $k^d$ new cubes which form $\Omega _1$.
On a cube-by-cube basis we continue recursively yet independently.

The construction process yields a tree of cubes which define a cascade of grids
$\Omega _0, \Omega _1, \ldots , \Omega _{\ell_{\text{max}}}$ that are
embedded into each other.
We call the subscript $\ell$ in $\Omega _\ell$ the grid
level\replaced[id=R1]{, i.e.~the smaller the index the coarser the mesh
\cite{Hackbusch:16:LargeSparseSystems,Trottenberg:00:Multigrid}, }{---this
differs to classic multigrid textbooks where the lowest level index is often the one with the finest mesh width---}and make a few observations:
The individual grids embed into coarser grids ($\Omega _{\ell -1} \subset \Omega
_\ell $).
Individual meshes $\Omega _\ell$ might cover only parts of the domain and might
yield disjoint submeshes.
The union of all meshes $\bigcup _\ell \Omega _\ell = \Omega _h$
is an adaptive mesh.
By subsequently removing the largest level $\ell $ from the union, we can
construct our coarse grid hierarchy in a geometric multigrid sense.
With this level definition, it is convenient to label $u_\ell $ with the
level instead of a generic $_h$ subscript.

\added[id=R1]{\emph{Correction vs.~full approximation storage realisations.}}
Multigrid solvers can be classified into
correction schemes and full approximation storage (FAS)
realisations\cite{Brandt:82:Guide,Trottenberg:00:Multigrid}.
The latter operate on a solution to the PDE on each and every level, whereas
in a classic correction scheme the coarse
grid weights have solely correcting semantics.
We make all inner and boundary points of each mesh $\Omega _\ell$ carry a
$d$-linear shape function.
$\bigcup _\ell \Omega _\ell = \Omega _h$ hence spans a generating system where
multiple vertices (and therefore weights) coincide
spatially yet are unique due to their level.
Let $P_\ell ^{\ell +1}$ denote prolongations of data on level $\ell $ onto level
$\ell +1$.
As there is no guarantee that all cubes of the mesh $\Omega _\ell$ 
are refined, the operator might
only take a subset of $\Omega _\ell$ and transfer it onto the next resolution
level.
Let $R_\ell ^{\ell -1}$ be a restriction, i.e.~the counterpart operation 
to $P_\ell ^{\ell +1}$.
Again, it affects only those regions of $\Omega _\ell$ that are refined further. 
There are three natural implications of this setup:
(i) We can make the overall solution to the PDE unique by 
enforcing $u_{\ell -1}(x) \gets u_{\ell }(x)$ for every vertex pair that
coincides spatially.
We write this down as $u_{\ell -1}(x) = I_\ell ^{\ell -1} u_{\ell }(x)$.
$I$ is the injection operator \cite{Griebel:90:HTMG} \added[id=R1]{(also named
``trivial restriction'' \cite{Hackbusch:16:LargeSparseSystems})}.
(ii) We can simply interpolate weights from $u_{\ell -1}$ to all
hanging vertices on level $\ell$.
Hanging vertices, i.e.~vertices with less than $2^d$ adjacent cells on the same
level, do not carry any real shape functions.
Yet, we temporarily augment them by truncated shapes such that a weak
Ritz-Galerkin formulation for their neighbouring non-hanging vertices on the
same level makes sense.
(iii) We can exploit the Ritz-Galerkin coarse grid operator definition and
implement Griebel's HTMG \cite{Griebel:90:HTMG} straightforwardly:
\ifthenelse{\boolean{specialissue}}{
\begin{eqnarray*}
 \underbrace{ A_{\ell -1} }_{=R_\ell ^{\ell -1}A_{\ell}P_{\ell -1}^\ell} 
   u_{\ell -1} 
  & = & A_{\ell -1} e_{\ell -1} + A_{\ell -1} I_\ell ^{\ell -1} u_{\ell}
  = R_{\ell} ^{\ell -1} 
  \underbrace{ r_{\ell} }_{=f_\ell - A_\ell u_\ell}
  + R_\ell ^{\ell -1}A_{\ell}P_{\ell -1}^\ell I_\ell ^{\ell -1}
  u_{\ell} = R_{\ell} ^{\ell -1} \left( f_{\ell} - A_{\ell}
  \underbrace{(u_{\ell} - P_{\ell-1}^\ell I_\ell ^{\ell -1} u_{\ell})}_{\hat
  u_\ell} \right)
\end{eqnarray*}
}{
\begin{eqnarray*}
 \underbrace{ A_{\ell -1} }_{=R_\ell ^{\ell -1}A_{\ell}P_{\ell -1}^\ell} 
   u_{\ell -1} 
  & = & A_{\ell -1} e_{\ell -1} + A_{\ell -1} I_\ell ^{\ell -1} u_{\ell}
  = R_{\ell} ^{\ell -1} 
  \underbrace{ r_{\ell} }_{=f_\ell - A_\ell u_\ell}
  + R_\ell ^{\ell -1}A_{\ell}P_{\ell -1}^\ell I_\ell ^{\ell -1}
  u_{\ell} 
  \\
  & = & R_{\ell} ^{\ell -1} \left( f_{\ell} - A_{\ell}
  \underbrace{(u_{\ell} - P_{\ell-1}^\ell I_\ell ^{\ell -1} u_{\ell})}_{\hat
  u_\ell} \right)
\end{eqnarray*}
}
This is an elegant rephrasing of FAS relying on two different types of
residuals: the standard residual $r_\ell = f_\ell - A_\ell u_\ell$ guides any
iterative update on a level, while its hierarchical counterpart $\hat r_\ell
= f_\ell - A_\ell \hat u_\ell$ feeds into its next coarser equation.
Both result from the same discretisation stencil.
FAS traditionally is introduced as a tool to handle non-linear equation systems.
It also pays off for in-situ visualisation as it holds the solution in a
multiscale representation and thus can allow users to zoom in and out.
It encodes levels of detail.
We use FAS as it makes the handling of hanging meshes simple
\cite{Reps:17:Complex,Weinzierl:18:Quasi}:
Fine grid vertices adjacent to a refinement transitions on the finest level
have two different semantics:
They carry correction and solution weights.
Both quantities work in different regimes---the solution weights converging
towards the ``real'' value while the corrections approach zero.
With FAS, we do not have to distinguish/classify them.
This argument rolls over recursively to all grid levels.

\added[id=R1]{\emph{Rediscretisation and geometric projections between levels
vs.~Ritz-Galerkin plus algebraic operators.}} 
Multigrid solvers can be
classified into solvers using rediscretisation plus
geometric transfer operators vs.~solvers using algebraic inter-grid transfer
plus Ritz-Galerkin correction operators\cite{Dendy:82:Black,deZeeuw90:MatrixDependent,Ruge:87:AMG}.
If we use geometric transfer operators, $P$ and $R$---we omit the
indices from hereon unless not obvious from the context---are
$d$-linear.
If we use rediscretisation, $A_\ell $ stems from a Ritz-Galerkin
\replaced[id=R1]{finite element}{Finite Element} discretisation with the same
type of shape and test functions on each and every level.
If we use algebraic inter-grid transfer operators, $P$ has to be constructed
such that any projection of a correction on level $\ell -1$ is locally mapped onto $A_\ell$'s
nullspace on level $\ell $.
We use BoxMG \cite{Dendy:82:Black,Dendy:10:Black,Yavneh:12:Nonsymmetric} to approximate such operators.
The restriction is either the transpose of $P$ or a symmetrised version of it
\cite{Yavneh:12:Nonsymmetric}.
A Ritz-Galerkin coarse grid operator $A_{\ell -1} = RA_{\ell }P$
is computed from the fine grid operator plus the (possibly algebraic) restriction and
prolongation.
Thus, it mimics the fine grid operator's
impact on a coarser solution. 
When using rediscretisation plus geometric inter-grid transfer operators one ``hopes'' that the
resulting coarse grid operator exhibits similar properties.

BoxMG plus Ritz-Galerkin yield the same operators as geometric rediscretisation
on our meshes if $\epsilon \in
\Omega$ is constant over the domain\replaced[id=R1]{ while}{.
In general,} it requires the absence of further PDE terms such as convection,
\deleted[id=R1]{the absence of} non-linear terms and
\replaced[id=R1]{anisotropic}{isotropic} material. \deleted[id=R1]{ that can be
modelled via a scalar $\epsilon $.
Such setups sound like toy problems.}
\replaced[id=R1]{The}{We emphasise however that the} finer the grid the more
dominant the second order term in most PDEs.
Furthermore, the finer the grid the ``smoother'' or rarer the $\epsilon$
transitions---unless we have totally random or noisy $\epsilon $ distributions.
As a result, there are often (fine grid) sections within our cascade of meshes
where geometric and algebraic operators are indeed the same or very close
\added[id=R1]{even though complex equation terms and material are present}.

\added[id=R1]{\emph{Multiplicative vs.~additive schemes.}
Multigrid solvers can be classified into additive solvers and
multiplicative ones~\cite{Bastian:98:Additive}.
Additive solvers determine the equation system's residual on the finest mesh
$\ell _{\text{max}}$, restrict this single residual to all levels $\ell $, and determine a
correction on all levels concurrently.
Finally these corrections are then added up, subject to a suitably defined
prolongations.
The simplest multiplicative solvers determine the residual on a level $\ell =
\ell_{\text{max}}$, smooth it, compute an updated residual, restrict this
residual to the right-hand side of the next coarser level, correct the solution there by applying the scheme recursively, 
and finally add all corrections subject to a prolongation to the fine grid
estimate.
}

\added[id=R1]{
Multiplicative solvers in general converge in fewer iterations
than their additive cousins, as an update on one level propagates through to other levels prior to updates on those levels.
A conceptional disadvantage of multiplicative multigrid is that solver steps on
the coarser mesh resolution levels tend to struggle to exploit all hardware
concurrency.
Cores start to idle.
Any strategy yielding tasks thus can expect that these tasks
exploit idling cores at one point.
This is one motivation for us to focus on improving concurrency for additive schemes---if the
approach works for additive solvers, it pays off for multiplicative
schemes too.
}

\subsection{adaFAC-x}


\deleted[id=R1]{Multigrid solvers can be classified into additive solvers and
multiplicative ones.
Additive solvers determine the equation system's residual on the finest mesh
$\ell _{\text{max}}$, restrict this single residual to all levels $\ell $, and determine a
correction on all levels concurrently.
Finally these corrections are then added up, subject to a suitably defined
prolongations.
Restriction, likewise prolongation, is often handled via $R \cdot R \cdot
\ldots \cdot R r_{\ell_{\text{max}}}$, i.e.~by applying a restriction
up to $\ell_{\text{max}}-\ell $ times.
The simplest multiplicative solvers determine the residual on a level $\ell =
\ell_{\text{max}}$, smooth it, compute an updated residual, restrict this
residual to the right-hand side of the next coarser level, correct the solution there by applying the scheme recursively, 
and finally add all corrections subject to a prolongation to the fine grid
estimate.
}

\deleted[id=R1]{
Multiplicative solvers in general converge in fewer iterations
than their additive cousins, as an update on one level propagates through to other levels prior to updates on those levels.
A conceptional disadvantage of multiplicative multigrid is that solver steps on
the coarser mesh resolution levels tend to struggle to exploit all hardware
concurrency.
Cores start to idle.
Any strategy yielding tasks thus can expect that these tasks
exploit idling cores at one point.
This is one motivation for us to focus on improving concurrency for additive schemes---if the
approach works for additive solvers, it pays off for multiplicative
schemes too.
}

Additive solvers are often used solely as preconditioner, as they yield inferior
convergence and face severe stability problems.
They tend to overcorrect the solution and this overcorrection is more
severe for increased number of grid levels \cite{Reps:17:Complex}.
It is thus convenient to damp the updates of the equation systems by a factor
$\omega $ that depends on the grid levels.
A convenient choice is to use an
$\omega $ on the finest level, $\omega ^2$ on the next coarser one and so
forth.
While this prevents overshooting, it tends to destroy multigrid convergence for
larger systems
as the elliptic nature of (\ref{introduction:equation}) implies that any local
change propagates through the whole domain.
With shape functions with local support, this propagation can only be realised
on coarser levels.
However, exponential damping with $\omega, \omega ^2, \omega ^3, \ldots$ and $ \
0<\omega<1$ effectively prevents changes from propagating rapidly via the coarse
resolutions.

An alternative to aggressive damping is additively damped
asynchronous FAC \cite{Murray:19:Dynamically}, which is a solver inspired by AFACx
\cite{Hart:89:FAC,Lee:04:AFAC,McCormick:86:FAC,McCormick:89:AFAC,McCormick:92:Multilevel}.
The term asynchronous refers to its additivity, i.e.~highlights that the
individual level updates can be computed without any synchronisation.
Additively damped denotes that a level's damping parameter
is similarly computed independently of all other updates as
an update for a single level is
\begin{equation}
  u_{\ell } \gets 
  \underbrace{
    u_{\ell } + S_{\ell}\left( R\hat r_{\ell +1} - A_{\ell}u_\ell  \right) 
  }_{\text{Standard additive scheme with FAS}}
  -
  \underbrace{ 
   \tilde P(S_{\ell - 1} \tilde R (R\hat r_{\ell +1}-A_{\ell}u_\ell))
  }_{\text{Additional additive damping of correction}}.  
  \label{equation:mg:adaFAC-x}
\end{equation}

\noindent
$S_\ell$ here denotes the smoother approximating the impact of $A_\ell^{-1}$.
Jacobi yields $S_\ell = \omega diag^{-1}(A_\ell)$, i.e.~here is where the 
original damping $\omega$ enters the equations.
The scheme (\ref{equation:mg:adaFAC-x}) \replaced[id=R1]{
projects a residual $r_\ell $ immediately one level
further---possibly using a modified restriction $\tilde R$---before it evaluates
the additive multigrid term on level $\ell $.
}
{
takes the solution of
additive multigrid on a level
$\ell $, and projects the arising $_\ell $ residual immediately one level
further, possibly using a modified restriction $\tilde R$.
}
When we update the solution on level $\ell $, we damp our actual update 
with the projection of an update step of this
further restricted equation.
The idea is that the auxiliary equation mimics the
overshooting potential of the real additive correction running concurrently
through the augmented $\tilde R$.
To achieve this, we either apply a smoothed prolongation operator $\tilde R$ to
our auxiliary equation---as we use a Jacobi smoother to construct such a
$\tilde R$, we speak of adaFAC-Jac---or choose $\tilde P, \tilde S$ and $\tilde
R$ such that the resulting smoother resembles a BPX-like scheme \cite{Murray:19:Dynamically}. 
The latter case uses solely $P$ and the injection $I$ for the auxiliary equations. 
We thus refer to it as adaFAC-PI.

adaFAC-x, i.e.~adaFAC-PI and adaFAC-Jac, are interesting additive multigrid
flavours, as they use Galerkin based operator constructions twice per level.
As a consequence, any inaccurate operator representation has twofold knock-on
effects on coarser levels.
With BoxMG and geometric operators, we have different combination opportunities
to construct our operators, and it is clear that the algebraic BoxMG variant is
subject to multifaceted input approximation inaccuracies if fine grid stencils
are not determined correctly.

\subsection{Numerical computation of stencils in a task language}


If all operators are geometric and we rely on rediscretisation on every level,
we can implement the multigrid scheme in a matrix-free way, once we embed the
entries of the vectors $u_\ell, r_\ell, \hat r_\ell, \ldots$  directly into the mesh
vertices.
Let our program traverse the mesh. 
Whenever we enter a cell, we load its adjacent $2^d$ vertices.
Each vertex holds its corresponding $u, r, \ldots$ entry, i.e.~one scalar per entry.
Hanging vertices hold interpolated weights or zero, respectively. 
As rediscretisation lacks dependencies on other operators we can construct all the
multigrid operator ingredients we need on-the-fly.
We compute them, apply them to the local data, and immediately add the impact
of the operator application (matrix-vector product) to the residual data within
the mesh.
Once all $2^d$ cells surrounding one vertex have been traversed, we can directly
apply our Jacobi point-smoother, restrict or prolongate all data
\cite{Weinzierl:Mehl,Reps:17:Complex,Weinzierl:18:Quasi}.
We traverse our grid cells and thus accumulate the matrix-vector impact
cell-wisely.
For this, we require cell-wise (element-wise) stiffness matrices.

Let $\mathcal{S}$ denote a smoothing task of the multigrid algorithm, i.e.~a
task that realises the update of $S_{\ell}$ from (\ref{equation:mg:adaFAC-x}) 
for one particular vertex/degree of freedom only.
Let $\mathcal{A}^{\text{(geo)}}$ denote a (geometric) assembly task
for a single element that is adjacent to the designated vertex/degree of freedom.
A geometric, matrix-free implementation of multigrid then issues a series of  
\begin{equation}
  \mathcal{S} \circ \underbrace{(\mathcal{A}^{\text{(geo)}} +
  \mathcal{A}^{\text{(geo)}} + \ldots +
  \mathcal{A}^{\text{(geo)}})}_{\text{$2^d$ assembly tasks for the $2^d$
  cells adjacent to one vertex}} 
  + \mathcal{S} \circ (\mathcal{A}^{\text{(geo)}} + \ldots +
  \mathcal{A}^{\text{(geo)}}) 
  + \mathcal{S} \circ (\mathcal{A}^{\text{(geo)}} + \ldots +
  \mathcal{A}^{\text{(geo)}}) 
  + \ldots
  \label{equation:matrix-free-geometric-multigrid-as-tasks}
\end{equation}
\noindent
tasks over the grid,
i.e.~unknown, entries.
All operators are to be parameterised over the levels.
As we work \replaced[id=R1]{element-wise}{element-wisely}, we do not evaluate
each cell task $\mathcal{A}^{\text{(geo)}}$ $2^d$ times.
Instead, we set up the element matrix once, and immediately feed its impact on
the surrounding $u_h$ values in the $2^d$ residuals.
The smoother $\mathcal{S}$ then acts on the residuals.
An assembly of the inter-grid transfer operators is omitted, as we know these
operators and can hard-code them.
The assembly determines one stencil, i.e.~integrates 
\begin{equation}
 \int _{\Omega _h} \epsilon \left( \nabla u, \nabla \phi \right) \ dx
  = 
 \sum _{c \in \Omega_h}
 \int _{c} \epsilon \left( \nabla u, \nabla \phi \right) \ dx.
 \label{equation:mg:weak-form-of-assembly}
\end{equation}
over $2^d$ cells $c$.
In our implementation, we exclusively work with an element-wise assembly where
$\mathcal{A}^{\text{(geo)}}$ computes the outcome of \ref{equation:mg:weak-form-of-assembly} over
one cell $c$.
Equation (\ref{equation:matrix-free-geometric-multigrid-as-tasks}) has
$2^d$ tasks that feed into one smoother though
each $\mathcal{A}^{\text{(geo)}}$ task which in turn feeds into $2^d$ $\mathcal{S}$ tasks---one
for each vertex that is adjacent to the cell. 
It is convenient to evaluate each cell operator once, feed it the $2^d$
follow-up steps, and thus to remove redundant tasks. 
We stress the entire procedure remains inherently additive however.

With explicit
assembly, we run 
\begin{equation}
 \left(
 \mathcal{S} \circ \ldots \circ \mathcal{S}
 \right) 
 \circ 
 \left( \ldots +
 \mathcal{A}^{\text{(geo)}}_{k^2h} + \mathcal{A}^{\text{(geo)}}_{kh}  + \mathcal{A}^{\text{(geo)}}_h 
 \right).
 \label{equation:level-operator-sequence-explicit-assembly} 
\end{equation}

\noindent
In (\ref{equation:level-operator-sequence-explicit-assembly}), the task symbol
$\mathcal{A}^{\text{(geo)}}_\ell $ is a supertask bundling the evaluation of all the
$\mathcal{A}^{\text{(geo)}}$ tasks on level $\ell $. 
This formalism relies on the insight that we can read multigrid cycles as
iterations over one large equation system comprising all coarse grid equations
if we commit to a generating system \cite{Griebel:90:HTMG}.
The addition in the assembly illustrates that submatrices within this large equation system can be
constructed (assembled) concurrently, as the individual cells on all levels are
independent of each other.

\begin{figure}

 \begin{center}
  \includegraphics[width=0.6\textwidth]{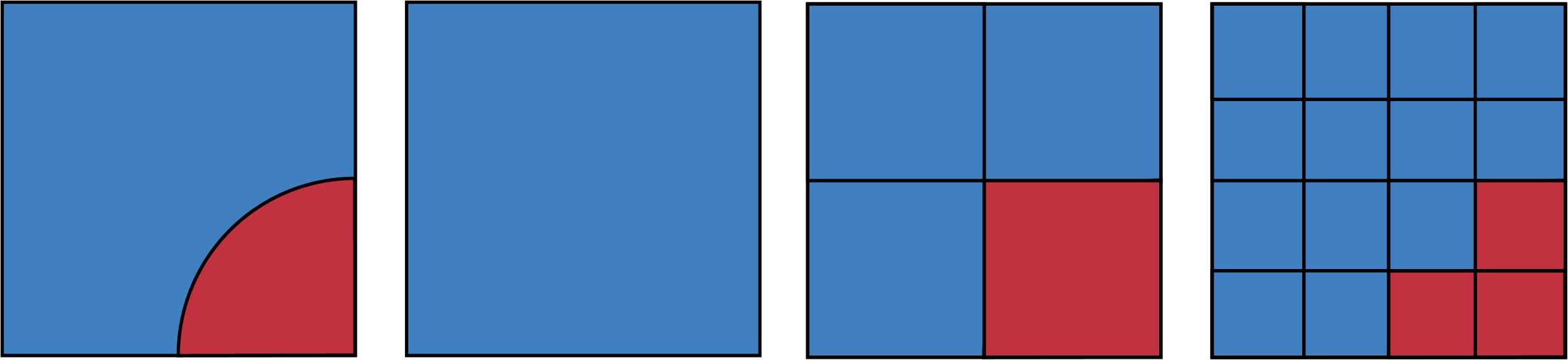}
 \end{center}
 \caption{
  Exact material parameter within a cell (left)
  and a splitting of the material parameter into $n^d$ quadrants for numerical integration (right).
  \label{figure:mg:material-illustration}
 }
\end{figure}

To make a \replaced[id=R1]{finite element}{Finite Element} discretisation
consistent, the assembly $\mathcal{A}^{\text{(geo)}}$ has to evaluate (\ref{equation:mg:weak-form-of-assembly}) over all
cells consistently, i.e.~in the same way for all of a cell's adjacent
vertices/stencils.
This is trivial for constant $\epsilon$, as we can
extract weights and $\epsilon $ from the integral and
\replaced[id=R1]{integrate}{solve the integration} over the remaining shape
functions analytically.
That is, if we know $\epsilon $ within a vertex, we can precompute
(\ref{equation:mg:weak-form-of-assembly}) for $\epsilon =1$ and scale it upon
demand.
If $\epsilon \not= const$, (\ref{equation:mg:weak-form-of-assembly})
the computation is less straightforward and typically has to be computed
numerically.
For this, one option is to approximate $\epsilon $.
A polynomial approximation makes limited sense, as we are particularly
interested in sharp $\epsilon $ transitions (material parameter jumps).
Higher order polynomials would induce oscillations.
We can however approximate $\epsilon$ as a series of constant values, i.e.~we
subdivide each cell into a Cartesian, equidistant subgrid with 
$n^d$ volumes.
Per volume, we assume $\epsilon $ to be constant.
For $n=1$, such a subcell integration is equivalent to sampling $\epsilon $ once
per cell centre (Fig.~\ref{figure:mg:material-illustration}).
The numerical integration can be expensive---not due to the arithmetic load
but due to the fact that $\epsilon $ lookups might be memory-access intense and
thus slow---which strengthens the case for an element-wise realisation of the
assembly, i.e.~it is better to make $\mathcal{A}^{\text{(geo)}}$ act per cell and feed into the
$2^d$ adjacent vertices rather than computing
(\ref{equation:mg:weak-form-of-assembly}) per $\mathcal{S}$ invocation.
The latter option would effectively integrate
(\ref{equation:mg:weak-form-of-assembly}) $2^d$ times.

As we know that different $\epsilon $ distributions require different choices of
$n$, it is convenient to parameterise the assembly tasks as
$\mathcal{A}^{\text{(geo)}}(n)$.
A fast assembly---either explicit or embedded into the solves---requires the
evaluation of $\mathcal{A}^{\text{(geo)}}(n)$ to be fast; in particular as we evaluate
each task once per cycle, i.e.~multiple times \replaced[id=R1]{overall}{per
complete solve}.
Therefore, it is in the interest of the user to choose $n$ as small as
possible---$\mathcal{A}^{\text{(geo)}}(n)$'s workload is in $\mathcal{O}(n^d)$---yet
still reasonably accurate.
Along the same lines, it is possible to distinguish different assembly
realisations by means of their accuracy, i.e.~whether we run them in double or
single precision.

\subsection{Stream-based matrix embedding into grid data structure}

\begin{figure}
 \begin{center}
  \includegraphics[width=0.5\textwidth]{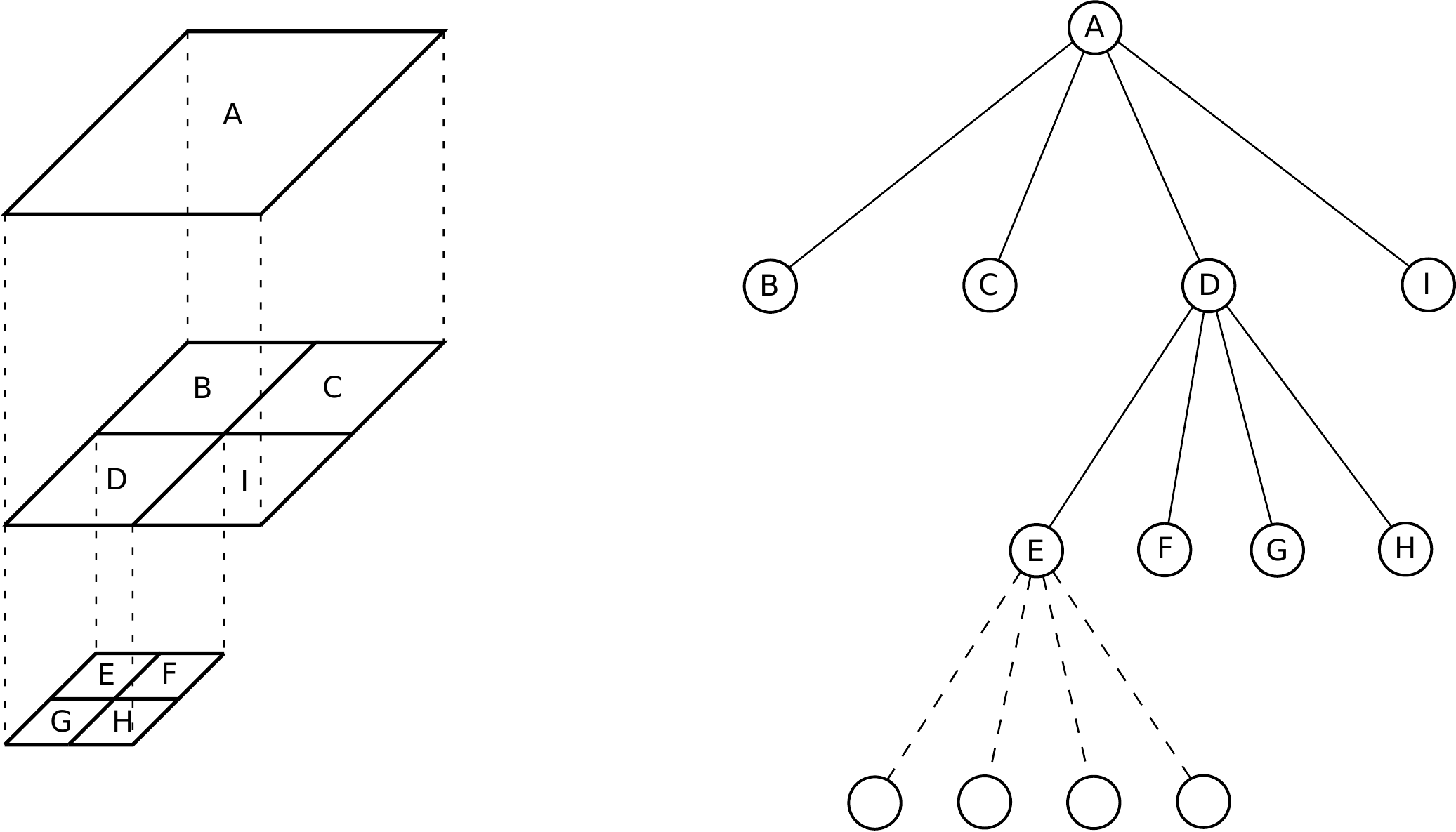}
 \end{center}
 \caption{
  \added[id=R1]{
   The tree nodes of a spacetree (right) span the multilevel adaptive Cartesian
   mesh (left) and also carry the local assembly matrices and $P$ and $R$
   operator parts.
   Fine grid nodes (B,C,E,F,G,H,I) carry geometric operators, while all others
   carry algebraic operators.
   If tree nodes refine (node E), then the operators
   of the refined node (E) have to switch from geometric to algebraic, which in
   turn requires recomputes of all coarser ancestors within the tree (A and D).
  }
  \label{figure:mg:tree-sketch}
 }
\end{figure}

An element-wise traversal of the set of meshes $\{ \Omega _0, \Omega _1, \ldots
, \Omega _{\ell_{\text{max}}} \}$ defines an ordering on the cells of the mesh.
It yields a stream of cells.
Additive multigrid's promise is that it exhibits a higher concurrency level than
multiplicative schemes.
Consequently, there's no order constraint on the cell enumeration within the
stream;
different to multiplicative multigrid where the pre-smoothing of cells within a level $\ell $ precede
cells of level $\ell -1$ creating causal dependencies.
It nevertheless pays off to realise such a partial level ordering, as it 
allows us to also integrate the bottom-up residual restriction within the
stream. 
If we run the grid traversal on a parallel machine, the cell stream is
split up into multiple streams.
Obviously, the cell ordering yields the exact data access pattern for the
vertices \cite{Weinzierl:Mehl,Weinzierl:19:Peano}, too, but this fact is not of
primary interest here.

As soon as permanent recomputation of the stencils becomes too expensive, we have
to memorise, i.e.~store the assembly matrices.
If we continue to avoid the maintenance of an explicit matrix data structure,
it is natural to store the local element matrices directly within the tree
cells.
We embed the stencils into the cell stream.
This avoids the memory overhead of matrix data structures requiring meta data,
sparsity pattern information and so forth, but it still has to pay for all
actual matrix entries.
They just are encoded within the mesh rather than within a dedicated data
container.
This approach works as long as we can guarantee that the cells are always
visited in the same order (per core/rank) by the traversal.
We work with our custom mesh and linear algebra implementation here.
Such a constraint however would allow the scheme to be realised within other
software offering matrix-free work through callbacks, too.


Once we have introduced this machinery, 
the in-stream storage can hold the algebraic operators
$\mathcal{A}^{\text{(alg)}}$ rather than their geometric counterparts:
on the finest mesh level, we make $\mathcal{A}^{\text{(alg)}}_{\ell
_{max}} = \mathcal{A}^{\text{(geo)}}_{\ell _{max}}$.
On all other levels, we make $\mathcal{A}^{\text{(alg)}}_{\ell
_{max}}$ hold the Ritz-Galerkin operator instead of a rediscretisation.
Finally, we can use the same implementation
technique to hold algebraic inter-grid transfer operators, too.
If all redundancies are eliminated---if we split up nodal operators over a
vertex's adjacent cells, the distribution of the stencil
is never unique---the memory footprint per cell within the stream thus grows by a
factor of $\frac{3^d + 2(2k-1)^d}{3^d}$.
We can now hold $P$ and $R$ explicitly within the stream however.

The present discussion introduces a storage scheme yet ignores the causal
dependencies in the computation of the stored data
\added[id=R1]{(Fig.~\ref{figure:mg:tree-sketch})}.
That is, we assume that all data held in-situ is readily available;
an assumption we explicitly pointed out to be wrong. 
A discussion on the computation rules plus the validity of all operators is
handled within the subsequent section.
Once we work with dynamically adaptive meshes, a task-based formalism inserts
grid update tasks $\mathcal{U}$ into the task graph.
These tasks either remove cells from the mesh or add cells.
In the former case, they also remove entries from the stream.
In the latter case, they insert entries.
In a multigrid environment with algebraic operators, 
$\mathcal{U}$ tasks change entries of
existing stencils:
removing cells switches Ritz-Galerkin correction operators into discretisation
stencils;
adding cells switches the held operators the other way round.
Both switches imply coarser operators changes that cascade, as the
Ritz-Galerkin condition ensures that coarser operators depend on finer ones.
This information update pattern is the subject of the discussion that follows.

\section{Stencil assembly variants}
\label{section:delayed-iterative}


To gain flexibility when and with what accuracy we assemble matrix entries, we
introduce per mesh cell $c$ a marker
\[
 p(c) = (p_1,p_2,p_3)(c) \in 
   \left\{ \mathbb{N}^+ \cup \{ \bot, \top \} \right\}
   \times
   \{ \bot, \top \}
   \times
   \{ 0,2,3,4,\ldots, 8 \}.
\]
We reiterate that we operate in a generating system/spacetree context, i.e.~there
are cells on each and every mesh level and different levels logically overlap.
The tuples are embedded into the cell stream as headers, i.e.~they precede the
matrix entries, and the tuple entries have the following semantics:

\begin{figure}
 \begin{center}
  \includegraphics[width=0.62\textwidth]{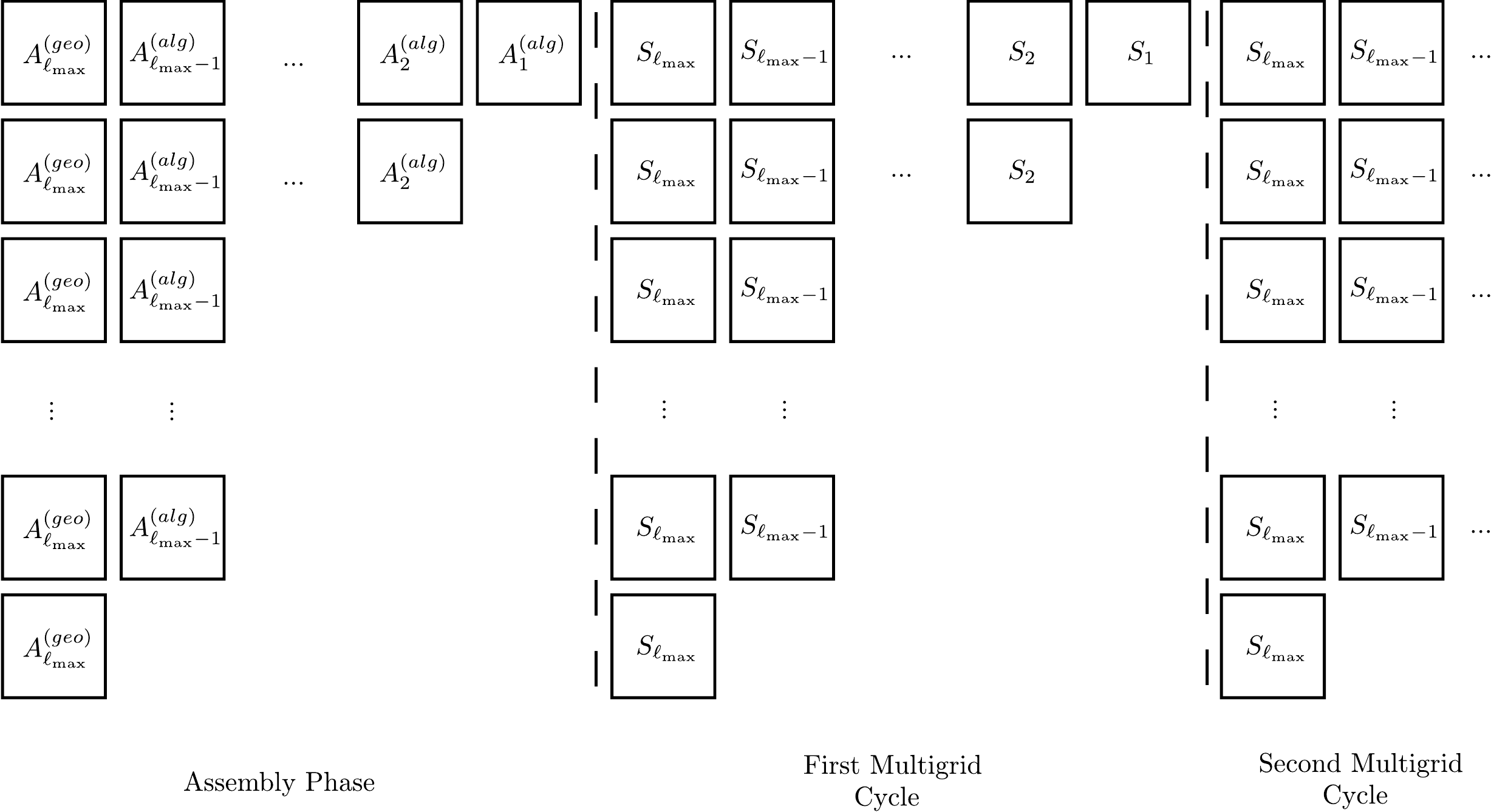}
 \end{center}
 \caption{
  \replaced[id=R1]{
   Schematic illustration of classic multigrid's task orchestration over six
   cores.
   Time runs from left to right.
   $A^{\text{(geo)}}_{\ell _\text{max}}$ denotes a stencil integration
   over a single cell on level $\ell _\text{max}$,
   $A^{\text{(alg)}}_{\ell }$ denotes an algebraic
   operator assembly (using $A^{\text{(alg)}}_{\ell +1}$), $S_\ell$ the
   application of one smoothing step on one cell, i.e.~the matrix-vector product feeding into
   the actual smoother. 
   The assembly feeds into the first smoothing step, i.e.~has terminated before
   we start cycling.
  }{ A representation
   of a task diagram of multigrid operations broken into tasks.
  The default sequential implementation (left)
  compared to our pipelined alternative (right).
  }
  \label{figure:intro:task-illustration}
 }
\end{figure}

\begin{figure}
 \begin{center}
  \includegraphics[width=0.45\textwidth]{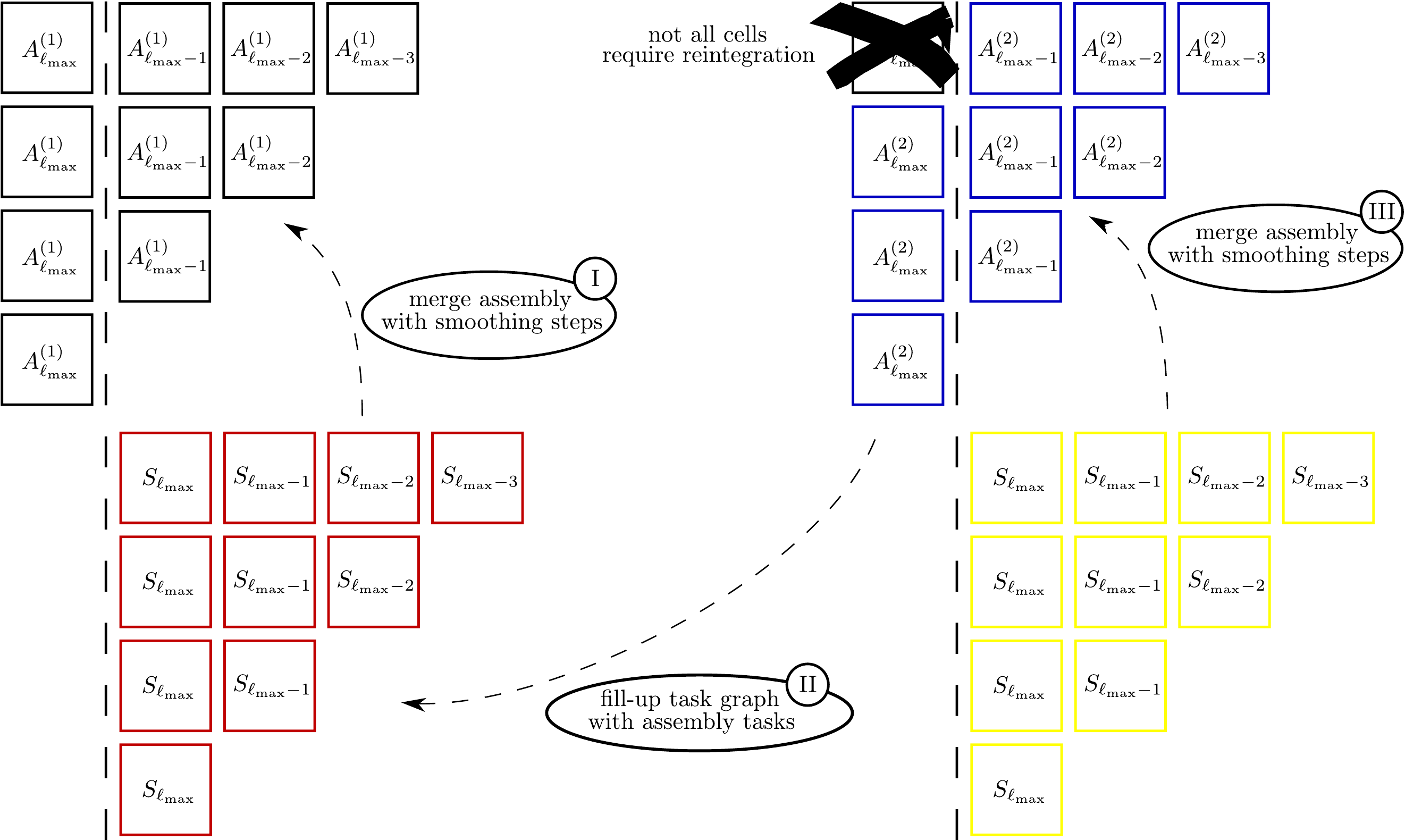}
  \hspace{2cm}
  \includegraphics[width=0.25\textwidth]{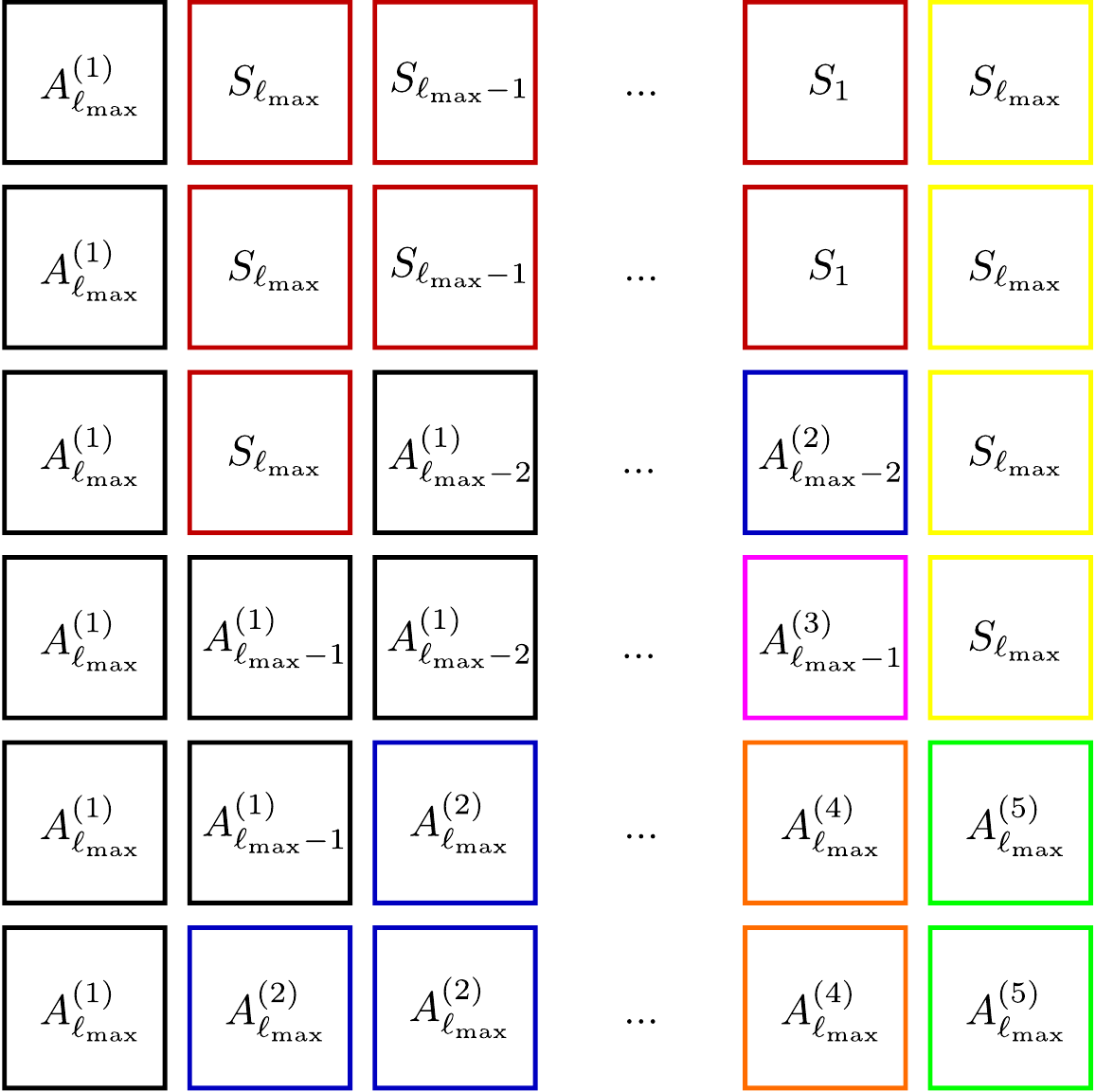}
 \end{center}
 \caption{
  \added[id=R1]{
   Left: Construction of our delayed assembly.
   The stencil integration $A_{\ell _\text{max}}$ is broken down into iterative
   substeps starting with a low-order approximation $A_{\ell
   _\text{max}}^{(1)}$. 
   We intermingle them with the first multigrid smoothing steps.
   Some stencils require further, more accurate integration $A_{\ell
   _\text{max}}^{(n)}$.
   Each stencil update requires us to recompute the algebraic coarse grid
   operators.
   Right: Example task orchestration/execution scheme on a six core system.
  }
  \label{figure:intro:pipelining-illustration}
 }
\end{figure}

\begin{itemize}
  \item[$p_1$] 
    If the first entry holds $\bot $, we have not yet performed any integration
    of (\ref{equation:mg:weak-form-of-assembly}) with any discretisation $n$.
    As a consequence, the matrix entry stream (matrix linearisation) does not
    hold entries for the corresponding stiffness matrix.
    If the first entry holds $\top$, our algorithm has concluded
    that this stencil is computed with sufficient accuracy already.
    The corresponding entries are found as the next $3^d$ or $3^d + 2(k-1)^d$
    entries within the stream.
    If the first entry holds a natural number $p_1(c) = n$, the matrix stream
    hosts the entries of the matrix, too.
    In this case, these entries result from an integration of
    (\ref{equation:mg:weak-form-of-assembly}) with an $n^d$ subgrid, and we
    cannot be sure yet that this $n$ is sufficiently large.
  \item[$p_2$] The second entry holds an atomic marker. 
    If it is set to $\top $, there is a task spawned into the task system which
    is currently computing a new approximation to the stencil, 
    i.e.~the code is in the process of evaluating
    (\ref{equation:mg:weak-form-of-assembly}).
    If the entry equals $\bot $ however,
    no update of the stencil is currently scheduled.
    The constraint $ p_1(c) = \top \Rightarrow p_2(c) = \bot $ holds.
  \item[$p_3$] For $p_1(c) = \bot$, the third entry has no semantics.
    Otherwise, it encodes in which precision the matrix entries are
    encoded.
    The marker determines how many bytes we have to continue to read within the
    stream to obtain our element-matrices and how these bytes have to be
    converted into floating-point numbers.
\end{itemize}


\subsection{Delayed stencil integration}

%
%
Let all cells initially carry $p_1(c)=\bot $.
The classic assembly phase
(\ref{equation:level-operator-sequence-explicit-assembly}) determines all
multigrid operators prior to the (first) assembly
\added[id=R1]{(Fig.~\ref{figure:intro:pipelining-illustration}, left)}, and
implicitly sets $\forall c: \ p(c)=(\top,\bot,8)$.
It is straightforward to implement a ``lazy'' implementation of the assembly along
the lines of lazy evaluation in functional programming languages\footnote{
 Our introductory paper sketching the delayed assembly strategy for the first
 time labelled the strategy as ``lazy''. It is however not lazy as it
 eventually computes all required matrix entries, and the term lazy furthermore
 has different semantics in programming languages. We thus dropped it in favour
 of ``delayed''.
}.
Here, lazy denotes an on-demand evaluation of functions just before their
result is required.
In the present case, this means that the local assembly matrix is computed just
prior to its first usage and then embedded into the stream for future use.
Given a fixed, global $n>0$, a lazy multigrid code tests in every cell whether
$p_1(c) = \bot$ before it runs the local assembly or matrix-vector product.
If $p_1(c) = \bot$, we either integrate (\ref{equation:mg:weak-form-of-assembly}) with an $n^d$ subgrid or compute
the multigrid stencils plus the inter-grid transfer operators due to the
BoxMG/Ritz-Galerkin formulation. 
Immediately after that $p(c) \gets (\top,\bot,8)$.

For multiplicative multigrid, this works naturally as we have a causal
dependency between levels.
We visit them from fine to coarse. 
Consequently, all level operators of level $\ell $ are available when we
hit $\ell -1$ for the first time. 
The coarse grid assembly $\mathcal{A}_{\ell -1} $ consisting of both the
construction of the element-wise operators plus inter-grid transfer
operator entries of $R_{\ell}^{\ell-1}$ and $P_{\ell-1}^{\ell}$ is thus by
definition already a series of ready tasks.
They have no incoming, unresolved dependencies.
They can be executed straight-away.
The observation holds for both geometric and algebraic multigrid
operator variants.

For additive multigrid, this straightforward \replaced[id=R1]{lazy}{delayed}
stencil integration works if and only if we stick to rediscretisation and geometric transfer operators.
It breaks down as we switch to algebraic operators, unless we prescribe the
order the levels are traversed, i.e.~unless we ensure that the traversal of
level $\ell $ is complete before we move to level $\ell -1$.
We may weaken this statement \cite{Weinzierl:Mehl,Reps:17:Complex} and
enforce that only those elements from level $\ell +1$ within the input of
a chosen vertex's local $P$ are ready.
While this might be convenient for many codes, it eliminates some of additive
multigrid's asynchronicity and thus one of its selling points.

\begin{observation}
 With a \replaced[id=R1]{lazy}{delayed} stencil integration, we can only utilise
 geometric \added[id=R1]{coarse-grid} operators in the first
 cycle\deleted[id=R1]{ if we do not impose further conventions}.
\end{observation}

\noindent
\added[id=R1]{
 Lazy evaluation is a particular flavour of a delayed operator assembly where we
 try to postpone the assembly as long as possible.
 If we weaken the information flow constraints within the assembly or the
 accuracy demands on the fine grid operator, we can construct further flavours
 of a delayed assembly: 
}

\subsection{Adaptive stencil integration}


If a proper global choice of $n$ for the fine grid (as well as for the
rediscretisation if we stick to geometric operators) is not known a priori, we can employ an
adaptive parameter selection:

Let $\mathcal{A}^{\text{(geo)}}(p)$ denote the assembly of the local assembly
matrix of one cell.
For the evaluation of (\ref{equation:mg:weak-form-of-assembly}), it is
parameterised by $p$, or $p$'s first tuple entry, respectively, i.e.~by the
numerical subsampling factor for the cell.
For an adaptive stencil integration, we make $\mathcal{A}^{\text{(geo)}}(p)$
accept the marker plus the current local element matrix as stored within the
stream if $p_1(c)\not = \bot$.
For $p_1(c)= \bot$, it is solely given the marker.  
\replaced[id=R1]{$\mathcal{A}^{\text{(geo)}}(p)$
returns}{$\mathcal{A}^{\text{(geo)}}(p)(p)$ yields} a new local element matrix
plus transfer operators, if required, plus an updated marker $p$ according to the following rules:

\begin{itemize}
  \setlength{\itemindent}{1.2cm}
  \item[$p_1(c)=\top$] The task returns immediately and we continue to work with
  the existing assembly matrix encoded in the cell stream.
  \item[$p_1(c)=\bot$] Assign $p_1(c) \gets 1 $ and evaluate
  (\ref{equation:mg:weak-form-of-assembly}) with one single sample in the centre
  of the cell. The resulting element matrix is pushed into the stream and hence
  used for subsequent calculations until this task is re-evaluated.
  \item[$p_1(c)\in \mathbb{N}^+$] The task evaluates
  (\ref{equation:mg:weak-form-of-assembly}) over the cell of interest. 
  It uses a $ (n+1)^d $ sub-grid with $n = p_1(c)$ to discretise
  $\epsilon $.
  The result is stored in a new element matrix $A^{\text{(new)}}$.
  Next, the task reads the previous matrix $A^{\text{(old)}}$ from the stream
  and compares the two matrices in the maximum element-wise matrix norm.
  This determines the new marker
  \begin{equation}
     p_1(c) \gets \left\{
    \begin{array}{ccc}
      \top & \text{if} & \frac{\| A^{\text{(new)}} - A^{\text{(old)}}
      \|_\text{max}}{\|A^{\text{(old)}} \|_\text{max}} < C
      \\
      n+1 & \text{if} & \frac{\| A^{\text{(new)}} - A^{\text{(old)}}
      \|_\text{max}}{\|A^{\text{(old)}} \|_\text{max}} \geq C
    \end{array}  
    \right.
    \label{equation:delayed-iterative:termination-criterion}
  \end{equation}
  Finally, we keep $A^{\text{(new)}}$ instead of $A^{\text{(old)}}$ within the
  stream, i.e.~\deleted[id=R1]{we }future work will utilise the former.
\end{itemize}

\noindent
It is obvious that a parameterisation of $P_\ell ^{\ell +1}$ and $R^\ell _{\ell
+1}$ makes limited sense.
However, BoxMG makes both operators depend directly on the operator on level $\ell
+1$.
This dependency propagates through all the way to the fine grid.
Therefore, both $P$ and $R$ depend indirectly on the $n$ choice of the
algorithm.

\begin{observation}
  After at most $n_{\text{max}} \cdot (\ell_{\text{max}}-1)$ steps
  ($n_{\text{max}} = \max _c \{ p_1(c) \}$), all local equation systems are
  valid, if the grid is stationary and we tackle a linear problem.
  $n_{\text{max}}$ \added[id=R1]{ is the maximum integration accuracy over all
  cells that is required eventually. It} is not known a priori.
\end{observation}

\noindent
The scheme describes an adaptive quadrature rule, where the accuracy of the
integrator is cell-dependent and determined by an iterative process.
This iterative process terminates as soon as a further increase of the accuracy
does not yield significantly improved stencils anymore.

\subsection{Asynchronous and anarchic stencil integration}


With the iterative scheme at hand, it is straightforward to construct an
asynchronous stencil integration, where the actual integration is deployed to a
task of its own and runs in parallel to the \replaced[id=R1]{solver's
iterations}{actual solve}
\added[id=R1]{(Fig.~\ref{figure:intro:pipelining-illustration}, right)}.
We augment our previously outlined adaptive integration such that
$p_2(c)=\bot$ holds initially for all cells.
Whenever we require a new stencil, we set $p_2(c)=\top$,
start the computation and set $p_2(c)=\bot$ upon completion.
In a synchronised setup, any check of $p_1(c)$ waits (blockingly) upon
$p_2(c)=\bot$.
$p_2(c)=\bot$ holds initially for all cells.
As long as $p_2(c)=\bot$, either no assembly task has ever been launched or the
task yielding the new, improved representation has already terminated. 
As long as $p_2(c)=\top$,  
we wait (and do meaningful other work in a reasonable task environment).

Alternatively, we can operate in an anarchic fashion and not wait upon the $p_2(c)$ entry.
Other than the initial stencil computation, 
we launch stencil tasks using 
\texttt{nowait} semantics, i.e.~we do not wait for this subtask to terminate.
\replaced[id=R1]{Each task}{These tasks} determines a new matrix 
$A^{\text{(new)}}$, \replaced[id=R1]{compares}{compare} $A^{\text{(new)}}$ to
the previously computed one, stores the new matrix, updates the $p$ marker and finally sets $p_2(c) \gets \bot$.
\replaced[id=R1]{It realises}{Thus realising} the third step of our adaptive
integration.
If the advanced integration of (\ref{equation:mg:weak-form-of-assembly}) 
has not terminated on time\replaced[id=R1]{,}{.} our solver continues to use
``old'' stencil approximations. 
\deleted[id=R1]{We rely on the fact that the result drops in eventually.}

\begin{observation}
 With an anarchic stencil integration, we have no control
 \deleted[id=R1]{which integration accuracy is used when. Furthermore,
 fine}{over when and with which integration accuracy we use. In either case,
 fine} grid stencil integrations are deployed to the
 background\replaced[id=R1]{ and once}{.
 Once} they drop in, \replaced[id=R1]{all affected}{the} coarse grid operators
 become invalid in a Ritz-Galerkin sense.
\end{observation}

\noindent
It is obvious that the iterative, delayed stencil integration can be
applied to all levels if we stick to rediscretisation.
With algebraic operators, the technique applies only to the finest grid level.
If we use iterative stencil integration on coarser levels, we have to control
the termination criterion in
(\ref{equation:delayed-iterative:termination-criterion}) carefully:
As the coarse equations are only correction equation and are ``only'' solved up
to a mesh-dependent accuracy in classic multigrid terminology, it makes limited
sense to choose the threshold $C$ there uniformly and small on all
resolution levels.

\subsection{Vertical Rippling}


If we work in a Ritz-Galerkin plus BoxMG environment, we can either deploy the
computation of the three arising coarse operators to background tasks, too, or we can
recompute these operators in each and every cycle.
Given the limited and deterministic computational load, the latter might be
reasonable.

\begin{figure}
 \begin{center}
  \includegraphics[width=0.4\textwidth]{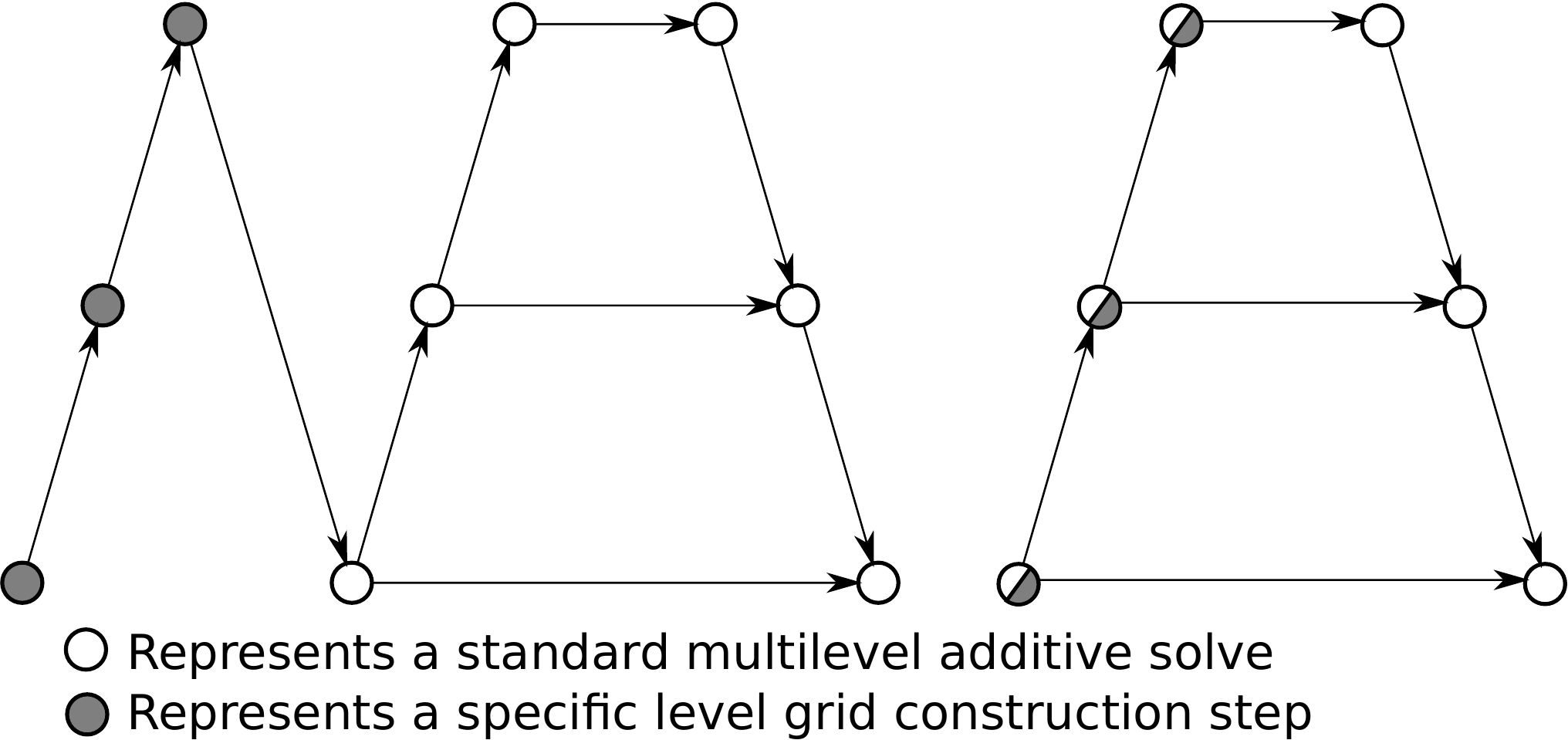}
 \end{center}
 \caption{
  Diagrammatic view of computing coarse grid equations  prior to a solver
  iteration (left) compared to plugging into a grid traversal of the actual solver.
  \label{figure:delayed:vcycle-illustration}
 }
\end{figure}

If the operators however are determined in the background, we can spawn only
those tasks that might actually yield changed operators.
An analysed tree grammar \cite{Knuth:90:AttributeGrammar} formalises the requirements:
If a matrix update changes the stencil associated with a vertex $v_\ell $ which
in turn is in the image of a stencil $P_{\ell -1}^{\ell}$ associated with a
vertex $v_{\ell -1}$, then the $2^d$ adjacent cells of $v_{\ell -1}$ should be
flagged.
In the next multigrid cycle, all flagged cells' $A_{\ell -1}, P, R$ computations
should be repeated, taking the new fine grid operator $A_{\ell }$ into account.
The same level-by-level information propagation---shown in 
Fig.~\ref{figure:delayed:vcycle-illustration}---formalises how information
propagates through both space and mesh resolutions if we recompute all three
operators in each and every multigrid cycle.

If we employ dynamically adaptive mesh refinement, the mesh coarsening or
refinement induces changes of operators. 
As a result, they implicitly trigger coarse grid operator updates.
A similar argument holds for non-linear setups: 
If the non-linear component induces signification changes in the fine grid operator---for
many operators, this might be a localised effect, i.e.~the fine grid equation
system might not change everywhere---we have to change a set of affected coarse
grid operators.
In both cases, it is important that we reset/remove $p_1(c)=\top$ from all
affected coarse grid cells.
Though $p_1(c) \gets 1$ obviously does the job, it might be convenient to
memorise the $p_1(c)\not= \top$ last used for a cell and to reset it to this
value instead.
In our implementation, we permanently recompute all coarse grid operators.

\begin{observation}
 In an additive setting, delayed operator updates ripple through the equation
 system\deleted[id=R1]{ iteratively}, i.e.~the updates propagate upwards by at
 most one grid level per cycle.
 From a certain point of view, coarse grid operators on a level $\ell $ lag behind the
 fine grid operators on level $\ell _{\text{max}}$ by $\ell
 _{\text{max}}-\ell $ iterations.
\end{observation}

\subsection{Precision toggling and information density}


As the local matrix accuracy increases,
the number of significant bits grows commensurately.
The number of bits
which carry actual information and not solely bit noise increases, too.
If a local stiffness matrix was computed using small $n$ values and
there is rapid variation in $\epsilon $ over the cell, 
then it makes limited sense to store the whole
matrix with eight bytes, i.e.~double precision.
Many bits represent accuracy that is not really there.
The same argument implies that a computation of
(\ref{equation:mg:weak-form-of-assembly}) can be initially done with reduced precision.
If a coarse correction operator is influenced by some cells where the integration has
not yet converged, it also makes limited sense to store the correction operator with
full double precision. 
$P, R$ and $A_{kh}$ could even be determined with half precision.
However, deducing a priori which accuracy is sufficient is a delicate task.
We propose a different approach.

Our code neglects the insight that we could use reduced-precision
computing---this might be unreasonable with new hardware generations support of half precision
(\texttt{binary16} or bfloats).
Instead we solely use double precision compute
routines, but focus on the potential of the significant bits w.r.t.~the memory
footprint:
In all of our previous descriptions, the third tuple entry of each cell is
set to $p_3(c)=8$.
This means that we use eight bytes, i.e.~double precision, to store each
datum.

Let the operators $A,P,R$ be stored as sequences of flexible precision values.
If $2 \leq p_3(c)<8$, each operator entry is encoded as follows:
The first bit is a sign bit.
The subsequent $8(p_3(c)-1)-1=m$ bits hold the mantissa.
The trailing eight bits hold the exponent.
Our exponent is a plain \texttt{signed char} which is chosen as close as
possible to the unbiased IEEE exponent.
The mantissa's bits are thus non-normalised.
If a double precision value is converted into this bespoke format, then we hold
it as
\[
 (-1)^{s} \cdot m \cdot 2^e \qquad 1\leq m <2 \ \text{stored in 52 bits
 ignoring the leading 1},
\]
i.e.~we encode the datum as $ (-1)^{s} \cdot \hat m \cdot 2^{\hat e}$
within the cell stream.
$\hat e = e$ if $-128 \leq e <127$.
Otherwise, $\hat e $ is either $-128$ or $127$, which is the closest value to $e$.
We ignore/remove the fact that $e$ is stored biased in the IEEE standard.
Our $\hat e$ is stored as C++ \texttt{char}, i.e.~without any shift. 
If $2 \leq p_3(c)<8$, we merely store the mantissa $\hat m$ as a truncated
representation as $m$:
we store its $p_3(c)-1$ bytes in the cell stream such that the total memory
footprint per matrix entry equals $p_3(c)$.

With this flexible storage format, we can compute the maximum error $\delta$ induced
by the new format after we have computed a new stencil/matrix:
Prior to any computation, we load the involved matrices from the cell stream and
convert them to IEEE double precision.
\replaced[id=R1]{If}{We run our algorithm if} an integration tasks computes a
new assembly matrix or if the Ritz-Galerkin/BoxMG routines determine a new
matrix, the matrices are \deleted[id=R1]{then} converted into a representation
where each matrix entry consumes only $p_3(c)$ bytes \added[id=R1]{prior to
their use}.

Throughout the conversion we determine the maximum error $\delta$ introduced by
our lossy compression.
In line with the increase of the approximation accuracy $n$, we also increase
the storage accuracy:
If one byte would have made a substantial difference, then we use a higher
precision.
Otherwise, we stick with the same precision.
For this, we usually employ two magic thresholds: one threshold guides the
adoption of $n$, one controls the increase of the storage footprint.
Both are independent of each other yet used within the same rules.

\begin{observation}
 Our modified storage scheme increases the number of bytes spent on the matrices
 where it is necessary. In subdomains and/or grid levels where high accuracy is
 not required, i.e.~where the operators lack detail, we use a low-precision data
 format.
\end{observation}

\noindent
It is clear that this approach is of limited value in its plain version.
However, it becomes powerful once we apply this idea not to the operators but to
the hierarchical surplus \cite{Weinzierl:18:Quasi}.
Let $A(n=1)$ be a local assembly matrix of a cell with one sampling point for
$\epsilon $ and $P^{(\text{geom})},R^{(\text{geom})}$ the operators resulting from $d$-linear
interpolation and restriction.
$P^{(\text{geom})}$ and $R^{(\text{geom})}$ can be hardcoded, while 
$A(n=1)$ is cheap to compute.
Instead of storing $A(n)$ or the real $P$ and $R$, we store only their
hierarchical surplus 
$\hat A = A(n) - A(n=1)$, $\hat P = P-P^{(\text{geom})}$, $\hat R = R-R^{(\text{geom})}$.
These hierarchical values typically are very small, i.e.~the resulting error
from the compression is very small, too. 
Most of the cells do not require eight bytes per matrix entry to encode their
(hierarchical) operators.

\begin{figure}
 \begin{center}
  \includegraphics[width=0.7\textwidth]{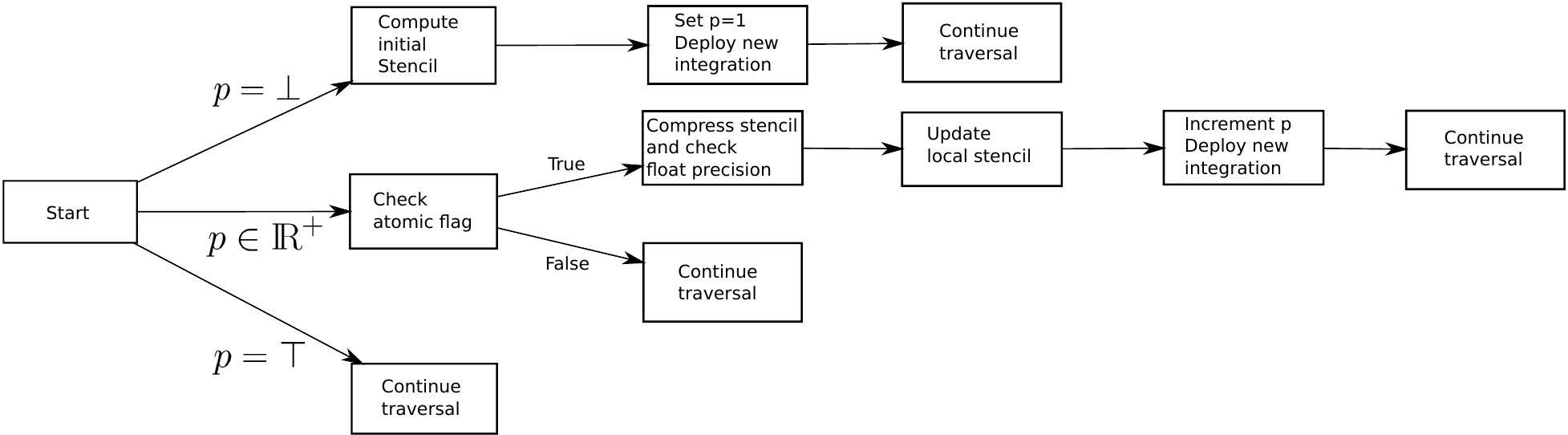}
 \end{center}
 \caption{
  	Flow chart of \replaced[id=R1]{stencil integration logic whenever we
  	enter a cell throughout the traversal.}{choice made upon encountering cell in
  	traversal}
  \label{figure:delayed:flow-chart}
 }
\end{figure}

This hierarchical representation finally motivates us to allow for entries
$p_3(c)=0$.
Whenever the hierarchical surplus of an operator
equals $n=1$ integration or $d$-linear transfer operators, respectively, it
holds a value close to zero.
If this value under the Frobenius norm is smaller than the prescribed threshold,
our converted, truncated format would encode a zero.
In this case,  we set $p_3(c)=0$ and skip all storage within the cell stream.
We need one byte for $p$ only.
This entire process is illustrated in Fig.~\ref{figure:delayed:flow-chart}.

\section{Results}
\label{section:results}


All experiments are run on an Intel Xeon E5-2650V4 (Broadwell)
with 12 cores per socket clocked at 2.4~GHz.
As we have two \replaced[id=R1]{sockets}{socket} per node, a total 24 cores
per node is available.
These cores share 64~GB TruDDR4 memory.
Shared memory parallelisation is
achieved through Intel's Threading Building Blocks (TBB),
though we wrap up TBB
with a custom priority layer such that we have very fine-granular control
over which tasks are run when.

\begin{figure}
 \begin{center}
  \includegraphics[width=0.4\textwidth]{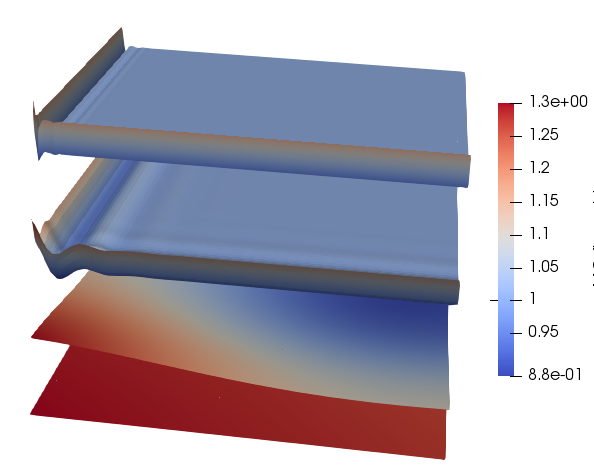}
  \includegraphics[width=0.3\textwidth]{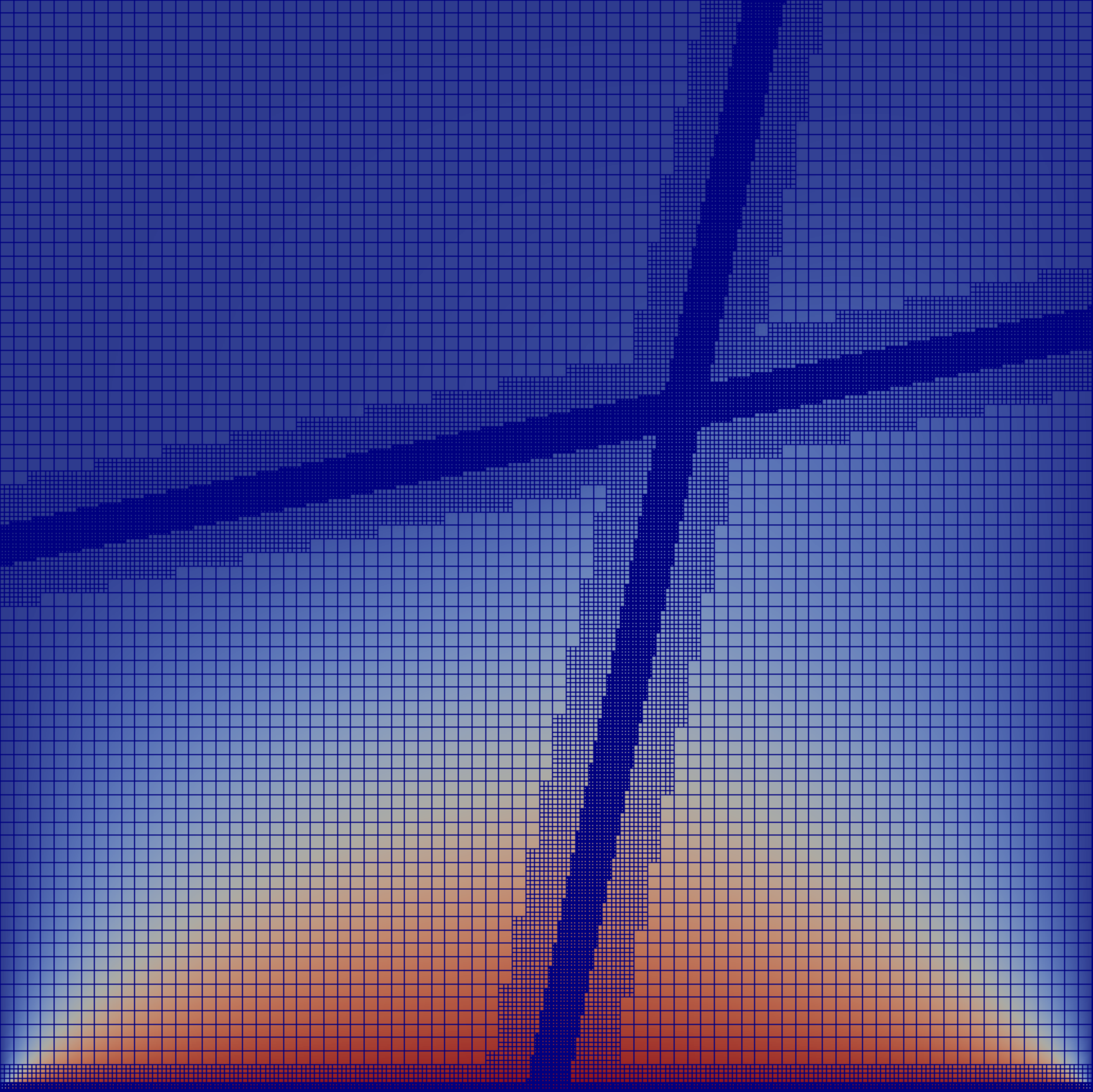}
 \end{center}
 \caption{
  Left: Illustration of $\epsilon $ for $\theta \in
  \{ 1/8,1,8,64 \}$.
  The smallest $\theta $ value is displayed at the bottom, the biggest on the
  top.
  Right: Adaptive mesh for our two-parameter setup where the discontinuous
  material transition in $\epsilon $ is not mesh-aligned.
  \label{figure:results:epsilon-illustration}
 }
\end{figure}


For the experiments, we stick to 
\[
  - \nabla ( \epsilon \cdot \nabla ) u = f
\]
on the unit square and supplement it with homogeneous Dirichlet conditions.
The setup is initialised with noise\replaced[id=R1]{ from $[0,\frac{4}{3}]$}{,
i.e.~unknowns are initialised with $u_h \in [0,\frac{4}{3}]$}.

We use two test setups. 
In the first one, we solve
\begin{equation}
 \qquad \text{with }
  \epsilon (x) = 
  1 + \frac{0.3}{d} \Pi _{i=1}^d e^{-\theta x_i} \cos( \pi \theta
  x_i).
  \label{equation:results:theta-setup}
\end{equation}
The above setup is ``simple'' as the analytic solution is $u(x)=0$ for $f=0$.
Yet, the parameter $\theta \geq 0$ allows us to play around with various
localised $\epsilon $ changes (Fig.~\ref{figure:results:epsilon-illustration}).
$\theta \gg 0 $ induces anisotropic behaviour close to the coordinate axes
and we thus know that our coarse grid operators struggle to capture the
solution's behaviour and thus degenerate.
Furthermore, inaccurate fine grid stencils here yield wrong results and the
bigger $\theta $ \added{is }the \replaced{greater}{higher} the number of integration points $n$ per cell \replaced{is required}{has to be chosen}.
However, the bigger $\theta $ \added{is }the more localised the significant $\epsilon $
changes\added{ are}, i.e.~non-uniform $n$ choices are mandatory.
Finally, additive multigrid tends to overshoot significantly if $\theta $ makes
$\epsilon $ change with high frequency.
Therefore we can readily expose any instabilities.

In the second setup, we employ only two $\epsilon $ values, and we keep them
constant over four regions of the domain
(Fig.~\ref{figure:results:epsilon-illustration}).
In two opposing regions, we choose $\epsilon = 1$.
The other two regions host $\epsilon \in \{10^{-1}, 10^{-2}, . . . , 10^{-5}\}$.
This setup is challenging as the jump between the material parameters can be
arbitrarily large and transition boundaries are not aligned with the
adaptive Cartesian mesh.
We cannot rely on the mesh to resolve the material changes properly.
We have to rely on the local assembly matrices to accommodate the $\epsilon $
layout.
As a consequence, a value of $n=1$ within the four subdomains is sufficient.
We might temporarily use $n=2$ to decide to switch $p_1(c)\gets \top$.
Where $\epsilon $ changes however, we easily obtain values of $n \approx 20$
before our algorithm decides that the approximation quality of the jump
is finally good enough.

All data report normalised residuals.
We measure the residual and divide it by the residual in the very first cycle. 
The solver stops as soon as the initial residual is reduced by ten orders of
magnitude.
Our data usually reports the number of fine grid solution updates also called
degree of freedom (DoF) updates.
Overhead workload due to coarse grids is not taken into account on any axis.

\subsection{
 Consistency with dynamic termination criteria and starvation effects
}


%
%
We kick our experiments off with some studies on dynamic termination
criteria\deleted[id=R1]{of the actual solve}.
Most codes terminate the solve as soon as the normalised residual runs under a given
threshold or stagnates.
If we use delayed, asynchronous stencil integration, i.e.~we do not wait per
cell for the underlying next step of the integration to terminate, we thus run
into the risk that we terminate the solve prematurely, i.e.~before the right local
assembly matrices have been computed.
From an assembly point of view, this is a starvation effect: The assembly tasks
are issued yet have not been scheduled and thus cannot affect the solve.
We end up with the solution to a ``wrong'' problem described by these inaccurate
operators.

%
%
We investigate this hypothesis simulating our test equation with a high
parameter variation on a regular Cartesian mesh hosting 59,049 degrees of
freedom.
Four multigrid \replaced{correction}{correct} levels are employed.
Our experiment tracks both the residual development and the number of background
tasks pending in the ready queue.
We reiterate that they are issued with low priority such that the incremental
improvement of the assembly process does not delay the
\replaced[id=R1]{solver iterations}{actual solve}.

\begin{figure}[htb]
 \begin{center}
  \includegraphics[width=0.45\textwidth]{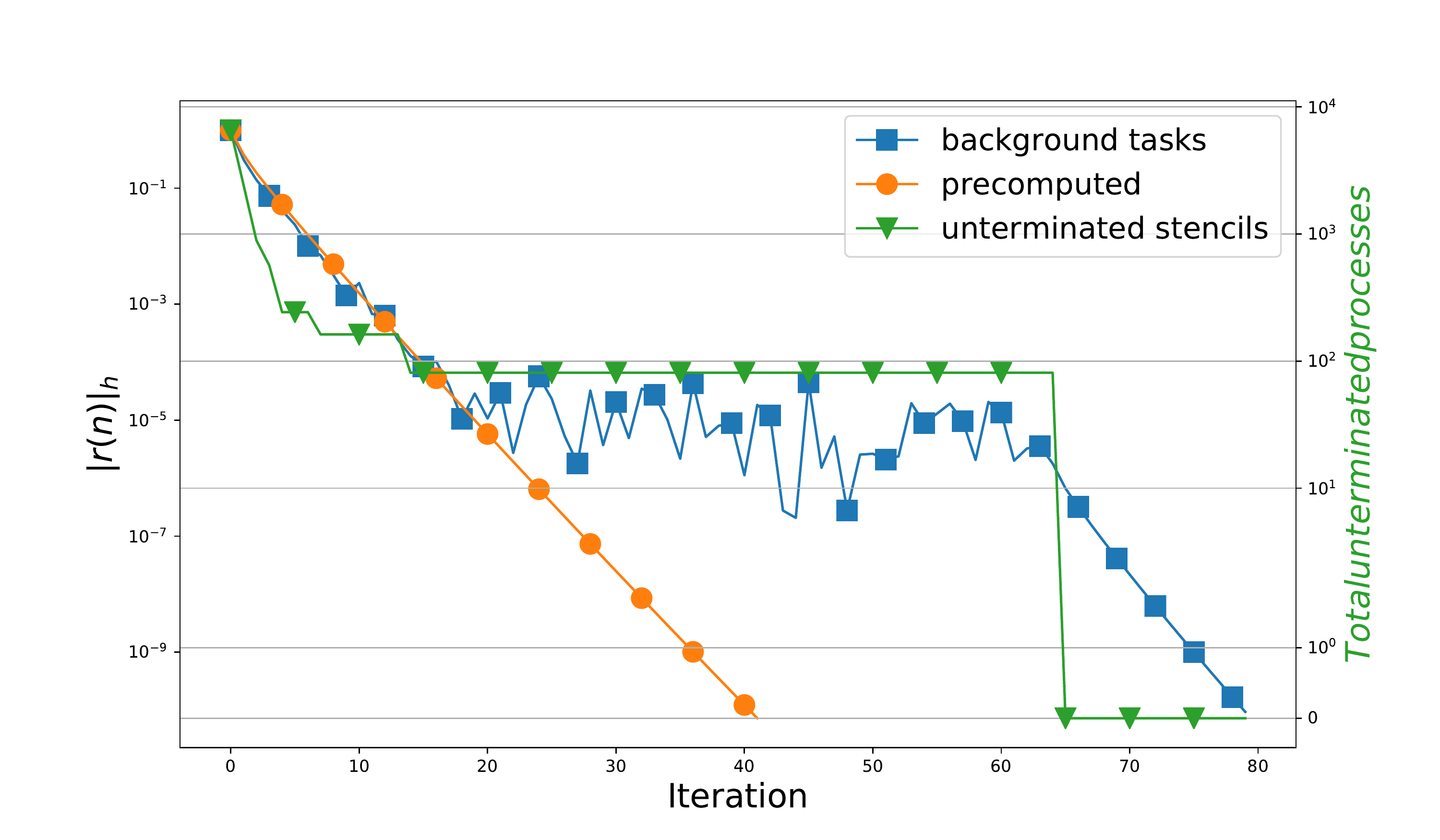}
  \includegraphics[width=0.45\textwidth]{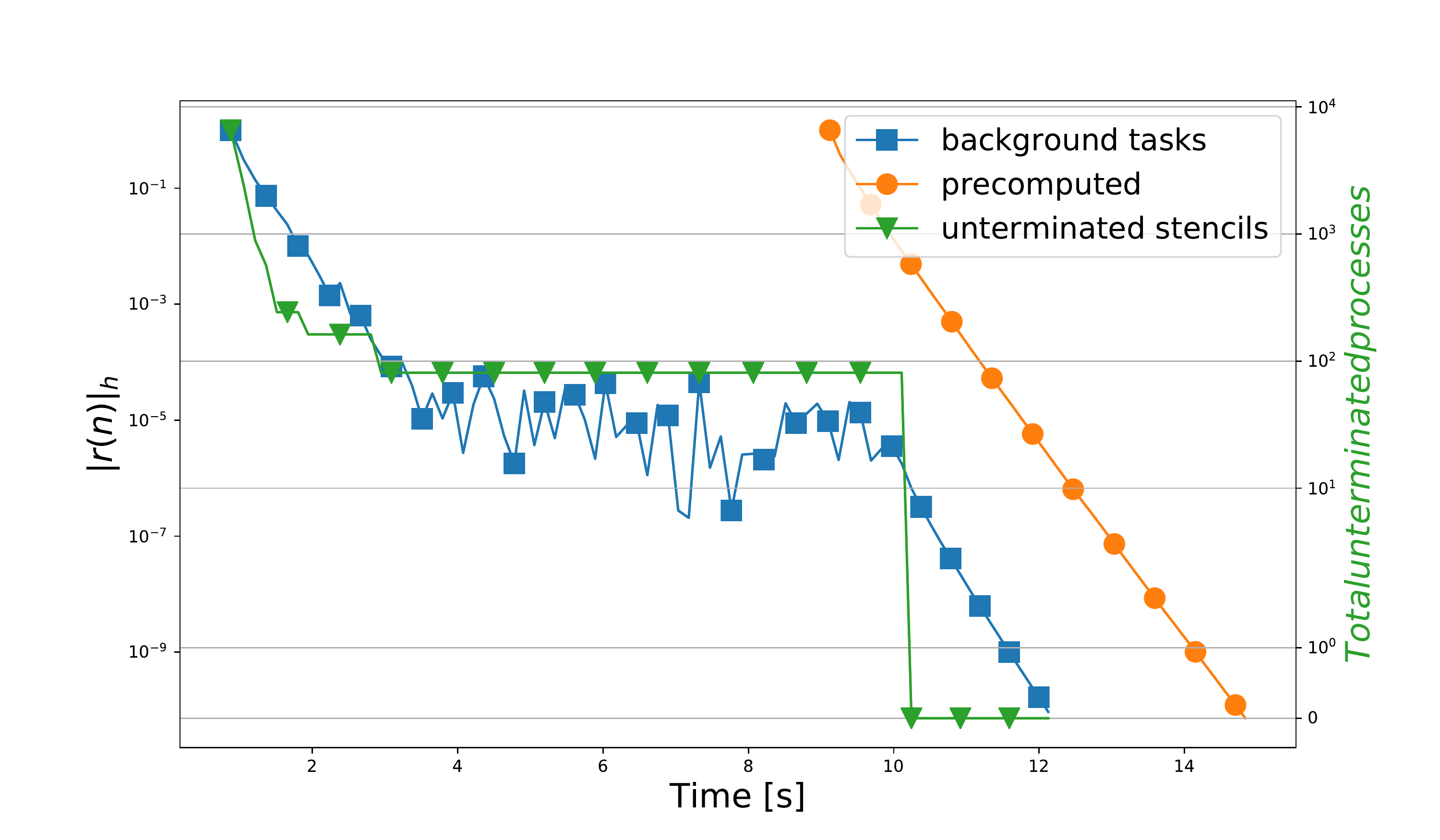}
 \end{center}
 \caption{   
 Convergence of delayed operator evaluation vs.~precomputed stencils/operators
 per iteration (left) and against real time (right).
   \label{figure:starvation}
 }
 \vspace{-0.3cm}
\end{figure}


%
%
For smooth $\epsilon $ distributions, we have not been able to spot any
deterioration of the residual evolution due to the delayed stencil integration
(not shown).
For rapidly changing $\epsilon$\added[id=R1]{, e.g.~$\epsilon = 1$ or $\epsilon =
10^{-5}$,} the solver's behaviour changes dramatically however
(Fig.~\ref{figure:starvation}).
Using a delayed assembly (stencil computation) deployed to background tasks, 
the solver iteration count required to reduce the normalised residual to a factor of 
$10^{-10}$ doubles compared to a solve where all operators are accurately computed prior to
the solve kick-off.
Initially, both methods show a similar rate of convergence, however the delayed
solve soon enters a regime where its residual almost stagnates around $10^{-5}$.
Throughout the initial residual decay, the number of pending background tasks
reduces dramatically.
While the residual stagnates, the number of background
integration tasks remains constant, however.
Towards the end of the residual plateau, the number of background tasks drops to
zero and the solve recovers and exhibits multigrid performance again.

%
%
Our solver spans one background assembly task per fine grid cell initially and
continues to work with a geometric approximation to the local assembly matrix
from thereon.
Most of the assembly tasks are associated with cells covering smooth $\epsilon $
distributions.
They thus discover that the assembly approximation is sufficiently accurate
almost immediately, i.e.~after increasing $n$ once.
They terminate and do not reschedule any tasks for this particular cell.
Only the few tasks associated with regions close to the significant $\epsilon $
variations require repeated rescheduling while increasing $n$.
By the time only these rescheduled tasks remain, 
the lack of accurate subcell material representations for some cells becomes
detrimental to the rate of convergence.
We reach a point where the current solution accuracy is balanced with the error
of the stencils/assembly matrices that still have to be integrated properly.
Updates to the cell matrices hence ``introduce'' error---or rather expose errors in the solution
that the previously held stencil was unable to account for.
Due to the elliptic nature of the operator these errors spread through the 
entire domain.
The entire solver stalls.
At the point all the background integration tasks have converged, i.e.~do not
reschedule themselves anymore, we regain multigrid convergence as we finally
solve the correct system that no longer changes.

\begin{observation}
 Dynamic termination criteria for the equation system solver have to be designed
 carefully with delayed operator assembly, as the solver might converge or
 stagnate towards a wrong solution.
\end{observation}

\noindent
While our convergence considerations seem to not favour the delayed,
asynchronous assembly, making the comparison with regards to real time
changes the picture   
(Fig.~\ref{figure:starvation}).
An increased iteration count in the solver is negated due to the headstart the
delayed evaluation gives the solver.
The setup also highlights that 
precomputing accurate stencils can take a greater amount of time
than the solve itself.
Finally, we see that the time-to-solution of the delayed assembly is superior
compared to the explicit a priori assembly.

As we kick off with low-accuracy operators, we effectively merge the first
few multigrid cycles with the actual assembly process:
The point in time at which delayed evaluation has computed an accurate solution
representation is a similar point in time to that when the precomputed stencil
has computed an accurate stencil;
even though our precomputation routines employ a dynamic $n$ choice as well.
Therefore the delayed method can be seen as a way of computing
a reasonably accurate initial guess,
and the delayed assembly manages to maintain the lead from its headstart.

\begin{observation}
 A delayed operator integration pays off in time-to-solution for rough material
 parameters.
\end{observation}

\subsection{
 Rippling with dynamically adaptive meshes
}


We continue with experiments where the grid is no longer fixed.
This adds an additional level of complexity, as the used coarse grid operators
change both due to the delayed integration plus due to the information
rippling.
In a traditional AMR/multigrid setup, any change in the grid necessitates
a change in all ``coarser'' equations.
This introduces a recompute step per mesh refinement.
Our methodology hides the recomputation cost
behind the solve.
On the downside, information propagates at most one level per
cycle up within the resolution hierarchy.

%
%
It is not clear whether such a massive delay in the coarse grid assembly could
lead into stability problems or severe convergence penalties:
The coarse grids are no longer acting upon the same equation as the fine grids
all the time.
While this is an effect affecting the previous experiments, too, dynamic
adaptive mesh refinement also makes the semantics of assembly matrices change:
After each refinement, former fine grid discretisations suddenly become
Ritz-Galerkin correction operators.
With our tests, we investigate whether 
they still continue to push the solution in a direction that effectively
minimises the error.
Our setup initially starts as a regular Cartesian mesh hosting 
68 degrees of freedom with a single multigrid correction level.
This increases due to the refinement to the order of 250,000 degrees of
freedom and seven multigrid correction levels.

\begin{figure}[htb]
 \begin{center}
  \includegraphics[width=0.33\textwidth]{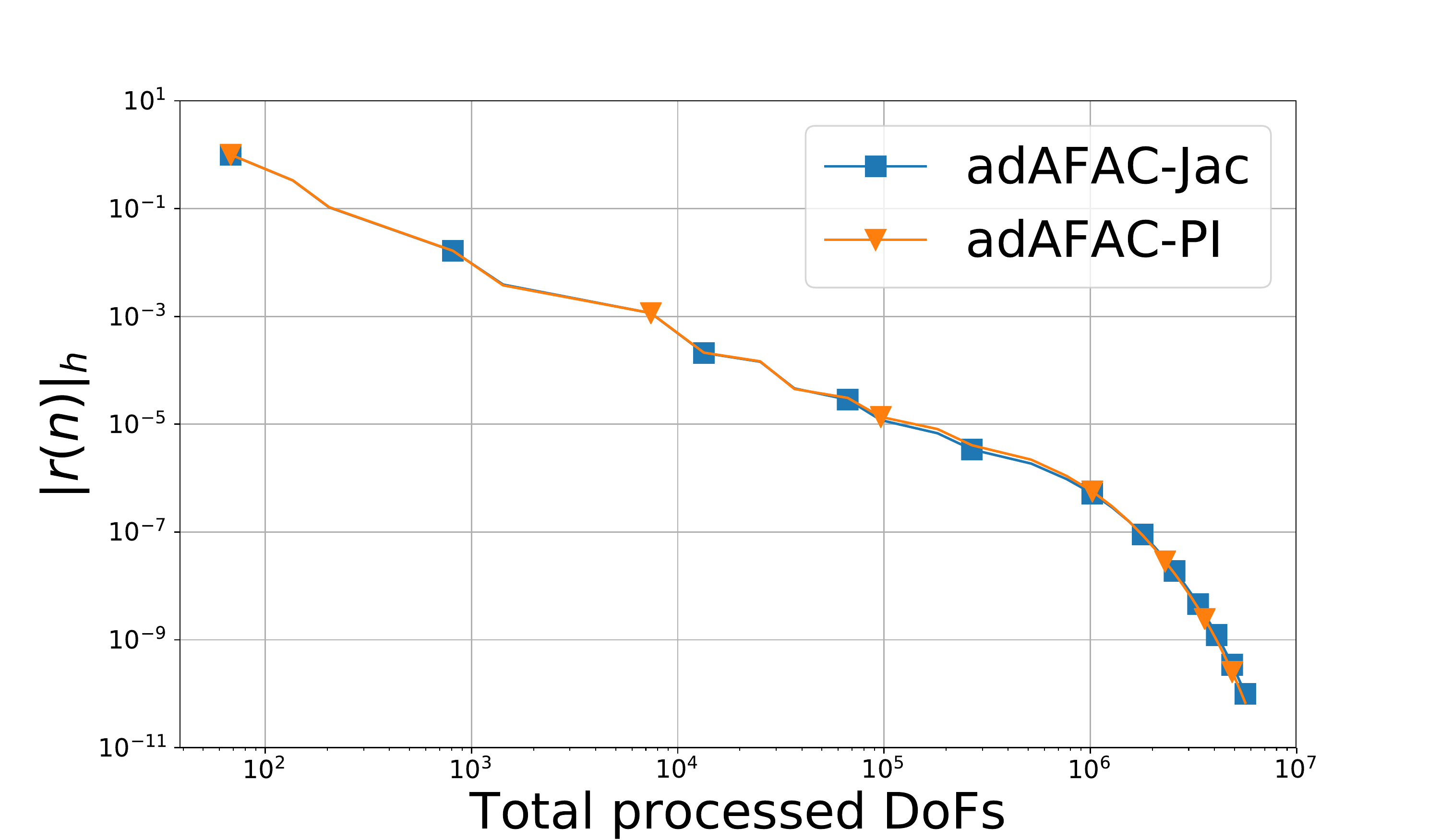}
  \includegraphics[width=0.33\textwidth]{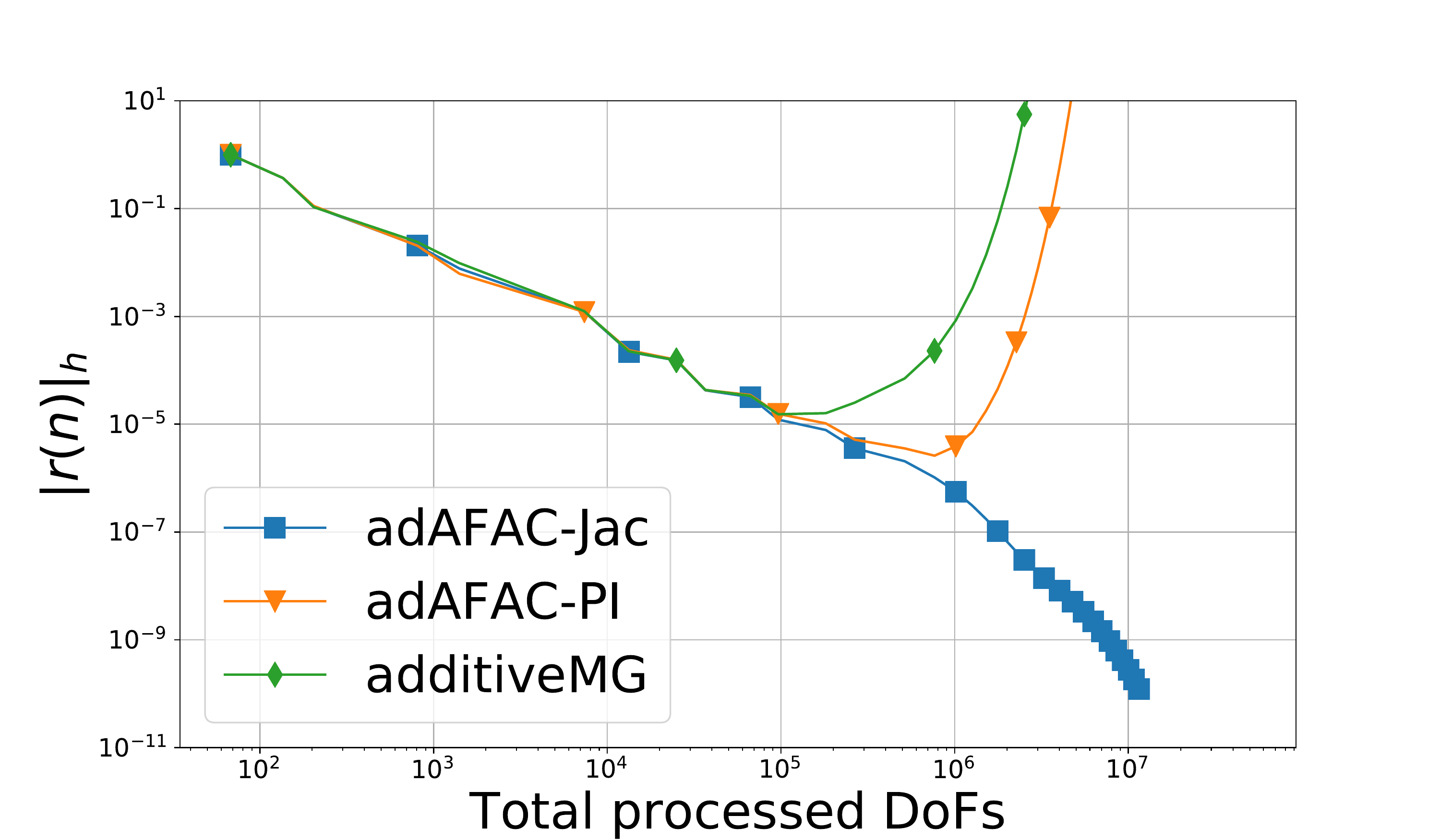}
  \includegraphics[width=0.33\textwidth]{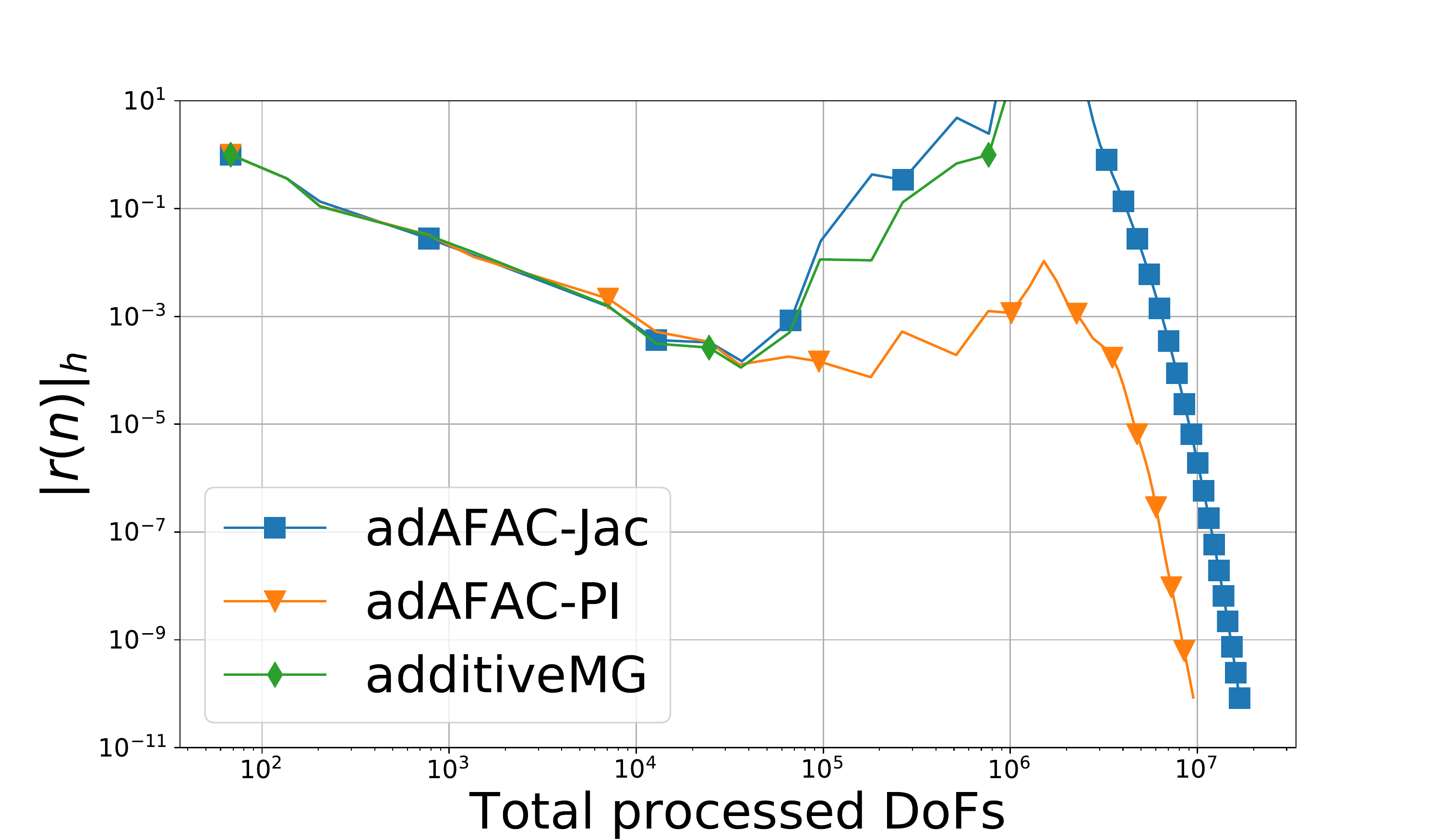}
 \end{center}
 \caption{
   Residual plots for the jumping coefficient problem and $\epsilon \in
   \{10^{-3},1\}$.
   All setups employ dynamically adaptive mesh refinement either reassembling
   all operators accurately (left) or using delayed operator assembly with
   geometric operators (middle) or algebraic operators (right).
   \label{figure:convergence:amr1}
 }
\end{figure}


%
%
We use our second test setup with only one type of a discontinuous material jump
over three orders of magnitude.
Initial results are given for $\epsilon \in \{1,10^{-3}\}$ 
(Fig.~\ref{figure:convergence:amr1}).
Our dynamic adaptivity criterion evaluates the solution's gradient over
$\Omega $ after each multigrid cycle and picks the degrees of freedom carrying
the top 10\% of the absolute gradient values.
We refine around these vertices and continue.
Convergence requires a total number of updates in the order of $10^7$ DoF
updates.
If a solve yields a residual that is 100 times bigger than the 
initial residual, we terminate the solver---even though the 
well-defined ellipticity implies that the solver eventually will ``converge
back''.
Our benchmark compares three different solver flavours with each other.
In the first variant, we immediately determine an accurate fine grid operator
whenever we refine the grid. 
Furthermore, we stop after each refinement and reassemble
all coarse operators accurately before we continue.
Algebraic BoxMG operators and Ritz-Galerkin are used. 
In the second variant, we continue immediately whenever we refine, but we use
geometric inter-grid transfer operators.
The fine grid operators are improved through delayed integration and eventually
yield improved Ritz-Galerkin coarse grid operators.
In the third variant, we finally let the inter-grid transfer operators ripple,
too.

%
%
The code with a complete re-assembly after each refinement step converges with a
rather shallow gradient first.
Throughout this phase, the grid is refined on alternate cycles. 
Once the grid becomes stationary, the solver exhibits a linear residual descent
with a steeper gradient.
If we combine geometric inter-grid transfer operators with dynamic refinement,
Ritz-Galerkin operator computations and delayed integration, the \replaced{adAFAC-PI}{adaFAC-PI}
solver variant suffers from a massive residual increase, while the adaFAC-Jac
variant outperforms our reassembly-based solvers by more than a factor of five.
If all operators are algebraic, this \replaced{adAFAC-Jac}{adaFAC-Jac} suffers from significant,
temporary residual explosions which eventually are recovered.
\replaced{adAFAC-Jac}{adaFAC-PI} is more robust yet still not as fast as its cousin with geometric inter-grid
transfer operators.

%
%
For both solver variants, our adaptive mesh refinement reduces the approximation
accuracy temporarily, as it replaces mesh cells likely fed by high accuracy
stencils with finer mesh cells with only one integration point.
This induces oscillations manifesting in temporary residual spikes.
As the fine grid cells start to improve their integration accuracy iteratively,
the overall system accuracy recovers.
Until this is complete, the residual can continue to increase by many orders of
magnitude, as the coarse grids solve an equation that is no longer a valid
correction and hence push the solution into the wrong direction.
With algebraic inter-grid transfer operators, this effect is more distinct than
with geometric operators: 
We know that geometric operators spanning big discontinuities induce
oscillations on the fine grid.
In the present case, we run into situations where algebraic inter-grid transfer
operators yield fine grid corrections anticipating the real material parameter
behaviour, but the new fine grid discretisation is not yet ready to mirror them.

\begin{observation}
 Rippling can cause dynamic mesh refinement to introduce massive residual
 deterioration.
\end{observation}

\noindent
Rippling yields temporarily incompatible equation system configurations.
A straightforward fix to this behaviour would be an inheritance mechanisms for
the number of cell integration points $n$:
If a cell results from a coarse grid cell with $n$ integration points, it could
immediately start a first fine grid integration with $\frac{n}{k^d}$
approximation points. 
However, such an approach would become a caricature of the delayed approach as
we would induce expensive assembly phases spread all over the solve.

\begin{figure}[htb]
 \begin{center}
  \includegraphics[width=0.65\textwidth]{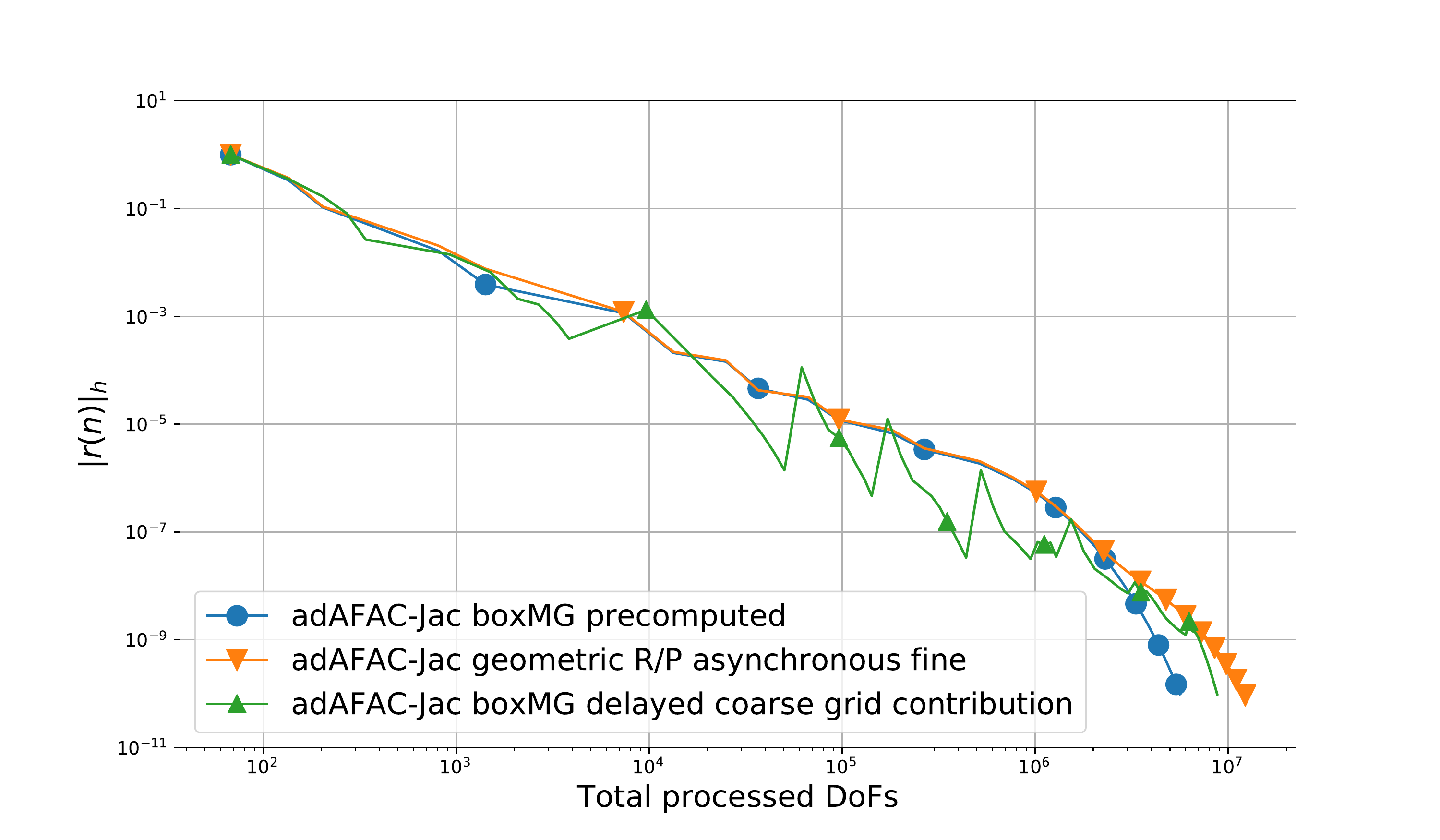}
 \end{center}
 \caption{   
   \replaced{adAFAC-Jac}{adaFAC-Jac} with $\epsilon \in \{1,10^{-5}\}$. We compare exact assembly to a
   Ritz-Galerkin coarse grid computation with geometric inter-grid transfer
   operators and a full algebraic formulation using BoxMG. 
   The latter two options temporarily switch off coarse grids until multigrid
   information has rippled through.
   \label{figure:convergence:amr2}
 }
 \vspace{-0.3cm}
\end{figure}


We continue with an even harder setup choosing  $\epsilon \in \{1,10^{-5}\}$
(Fig.~\ref{figure:convergence:amr2}).
However, we use the insight about the maximum rippling speed as follows:
We negate correction steps on a coarse level after each refinement until we
observe that the next finer level operator we depend on has been
updated at least once---either due to the Ritz-Galerkin recomputation or due to the delayed
integration.
This idea recursively applies to the next finer level if it is refined further.

We recover the stability of multigrid with an explicit reassembly though we need
around twice as many DoF updates compared to a classical version.
The improved stability is not dissimilar to classic multigrid theory where the
F-cycle requires a higher order interpolation. 
We use classic $d$-linear interpolation here whenever we introduce new vertices.
As we switch \replaced[id=R1]{off}{of} the coarse grid corrections, we
effectively smooth out this interpolation with a Jacobi step before we continue with multigrid.
The multigrid in turn is not switched on immediately but we effectively work our
way through a two-grid code, three-grid code, and so forth.
In the first solver phase where we add new grid elements frequently, we \added{only }run
series of fine grid smoothing steps \deleted{only }for the majority of the cycles.
The residual decays nevertheless, as most errors that can be resolved by newly
introduced vertices here are high frequency errors which are damped out
efficiently.
At the same time, switching off coarse grid corrections tends to free compute
resources which can be used to handle further stencil integrations.

\begin{observation}
 It is reasonable to pair up delayed stencil integration with a careful choice
 of which coarse grid operators are ready to be used in a multigrid cycle.
\end{observation}

\subsection{
 Memory footprint implications
}


%
%
For our memory footprint studies, we return to the first test setup and study
both a grid with $h \approx 0.004$ and one with $h \approx 0.038$.
They are small yet allow us to showcase the impact and behaviour of the proposed
techniques.
For different choices of $\omega $, we track the average number of bytes per
element matrix entry, the maximum number of integration points per cell that we
need, and the amount of memory saved due to delayed assembly in combination with our
hierarchical storage and data compression.
Our data focuses on the fine grid only, i.e.~we neglect coarse
grid effects.
They would blur the message and contribute to the memory footprint only
marginally.
The setup is configured such that the absolute error that we
introduce by storing a truncated version of the hierarchical surplus is at most $10^{-8}$.
This is close to machine precision.
As we work with the hierarchical surplus, it is reasonable to use an absolute
value rather than a relative value.
Finally, we use delayed operator approximation and stop the iterative
computation as soon as the relative difference between two subsequent
evaluations with $n$ and $n+1$ do not differ by more than one percent anymore.

\begin{table}
  \caption{
   We track the maximum number of integration points per axis $n$, the average
   number of points used over all cells, and the compression factor on a fine
   grid with $h \approx 0.038$ in (\ref{equation:results:theta-setup}).
   \label{table:memory-footprint}
  }
  \ifthenelse{\boolean{specialissue}}{}{
  \footnotesize
  }
  \begin{center}
    \begin{tabular}{l|rrr|rrr|rrr}
      & \multicolumn{3}{|c}{$\theta = 1$} 
      & \multicolumn{3}{|c}{$\theta = 16$} 
      & \multicolumn{3}{|c}{$\theta = 64$} 
      \\
      Cycle & max\{n\} & average $n$ & compression 
            & max\{n\} & average $n$ & compression 
            & max\{n\} & average $n$ & compression 
      \\
      \hline
      1 
       & 1 & 1.00 & 128.00
       & 1 & 1.00 & 128.00
       & 1 & 1.00 & 128.00
       \\
      2 
       & 1 & 1.00 & 128.00
       & 2 & 1.00 & 118.00
       & 2 & 1.07 & 22.64
       \\
      3 
       & 1 & 1.00 & 128.00
       & 3 & 1.00 & 118.00
       & 3 & 1.15 & 22.64
       \\
      4 
       & 1 & 1.00 & 128.00
       & 4 & 1.00 & 118.00
       & 4 & 1.22 & 22.64
       \\
      5 
       & 1 & 1.00 & 128.00
       & 5 & 1.01 & 118.00
       & 5 & 1.29 & 22.64
       \\
      6 
       & 1 & 1.00 & 128.00
       & 6 & 1.01 & 118.00
       & 6 & 1.36 & 22.64
       \\
      7 
       & 1 & 1.00 & 128.00
       & 7 & 1.01 & 118.00
       & 7 & 1.44 & 22.64
       \\
      8 
       & 1 & 1.00 & 128.00
       & 8 & 1.01 & 118.00
       & 8 & 1.51 & 22.64
       \\
      9 
       & 1 & 1.00 & 128.00
       & 9 & 1.01 & 118.00
       & 9 & 1.58 & 22.64
       \\
      10 
       & 1 & 1.00 & 128.00
       & 10 & 1.01 & 118.00
       & 10 & 1.65 & 22.64
       \\
      \hline
    \end{tabular}
  \end{center}
\end{table}

%
%
For a rather smooth parameter choice, we see that the delayed operator
integration stops after it has tested the local assembly matrices for $n=2$ 
against $n=1$ variants.
It cannot identify a difference exceeding one percent (Table
\ref{table:memory-footprint}).
The lowest \replaced[id=R1]{accuracy}{order} approximation is consequently used all the way through. 
We store the used assembly matrix relative to the $n=1$ approximation.
As they are the same here, we do not actually have to store the matrix, but it
is sufficient to bookmark a one-byte marker that flags that there is no
difference. 
A full element stiffness matrix requires $2^d \cdot 2^d \cdot 8$ double entries,
i.e.~128 bytes.
The marker can be held in one byte.
We compress the matrix data by a factor of 128.

If we use \replaced{the}{a} rougher $\epsilon $ distribution in
(\ref{equation:results:theta-setup}), the delayed stencil integration 
increases the approximation \replaced[id=R1]{accuracy $n$}{order} of at least
one cell per iteration.
\replaced[id=R1]{As this}{Both the fact that it} monotonously grows
\replaced[id=R1]{while}{and that} the average of the integration point
choices remains close to 1 suggest that this is an extremely localised effect.
Most of the cells \replaced[id=R1]{are sufficiently accurate with $n=1$, though}{seem to
be ``satisfied'' with rather low $n$ choices, but} the difference between the $n=1$
rediscretisation and the actual element matrices is not negligible anymore. 
We have to store the hierarchical difference and thus reduce the overall
compression factor.

For even higher $\theta $ choices, this reduction in compression efficiently
becomes more dominant.
The average $n$ also starts to grow, i.e.~more cells require
\replaced[id=R1]{a}{an} local assembly matrix which differs from a simple $n=1$
rediscretisation.
We still reduce the memory footprint by more than one order of magnitude
however.

\begin{observation}
  Our delayed integration in combination with compressed hierarchical storage
  makes the memory footprint of the solver increase over time. 
\end{observation}

\noindent
We ``overcompress'' all operators initially and gradually approach the 
most aggressive compression we can use without a loss of significant bits.
This is an advantageous property for algorithms with dynamic adaptivity which
build up the mesh iteratively.
As they start from small meshes, there is a low workload for the first few
iterations.
The required memory, i.e.~data amount to be transferred forth and back between
cores and main memory, increases as the solver continues and becomes more
expensive.
Conversely, a finer mesh width eradicates the $n$-distribution observations,
i.e.~the simulation can use $n=1$ all over the domain.
For $h \approx 0.004$, we have not been able to observe any increase in $n$ for
the three setups from above.
We return to a compression ratio of 128 and an
average $n$ \deleted[id=R1]{of} close to 1.

\begin{observation}
 If the total memory footprint increases due to dynamic AMR, the delayed
 integration and compression in return reduce the average memory footprint per
 mesh cell.
\end{observation}

\subsection{
 Scalability impact
}


\begin{figure}
 \begin{center} 
  \includegraphics[width=0.48\textwidth]{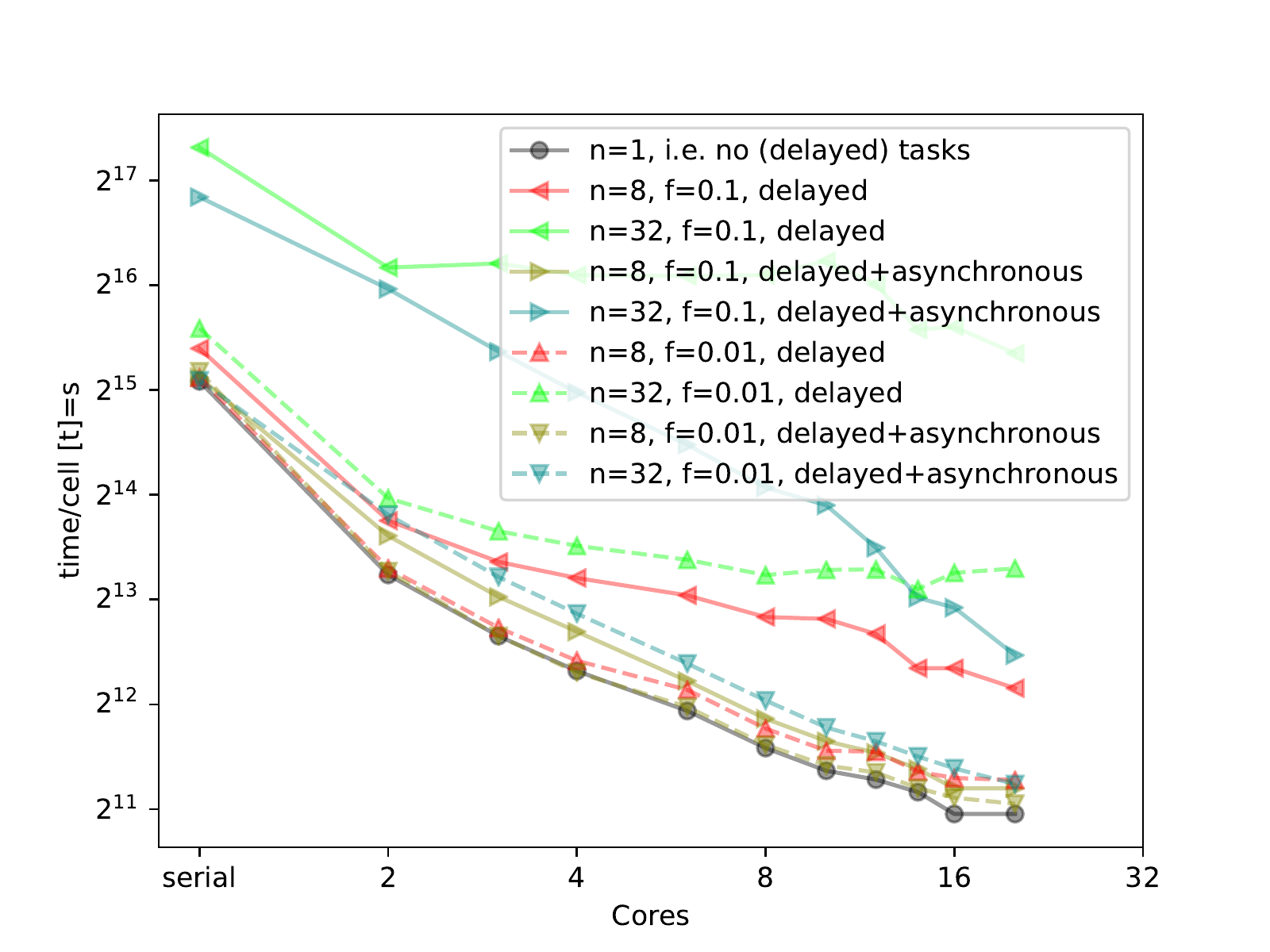}
  \includegraphics[width=0.48\textwidth]{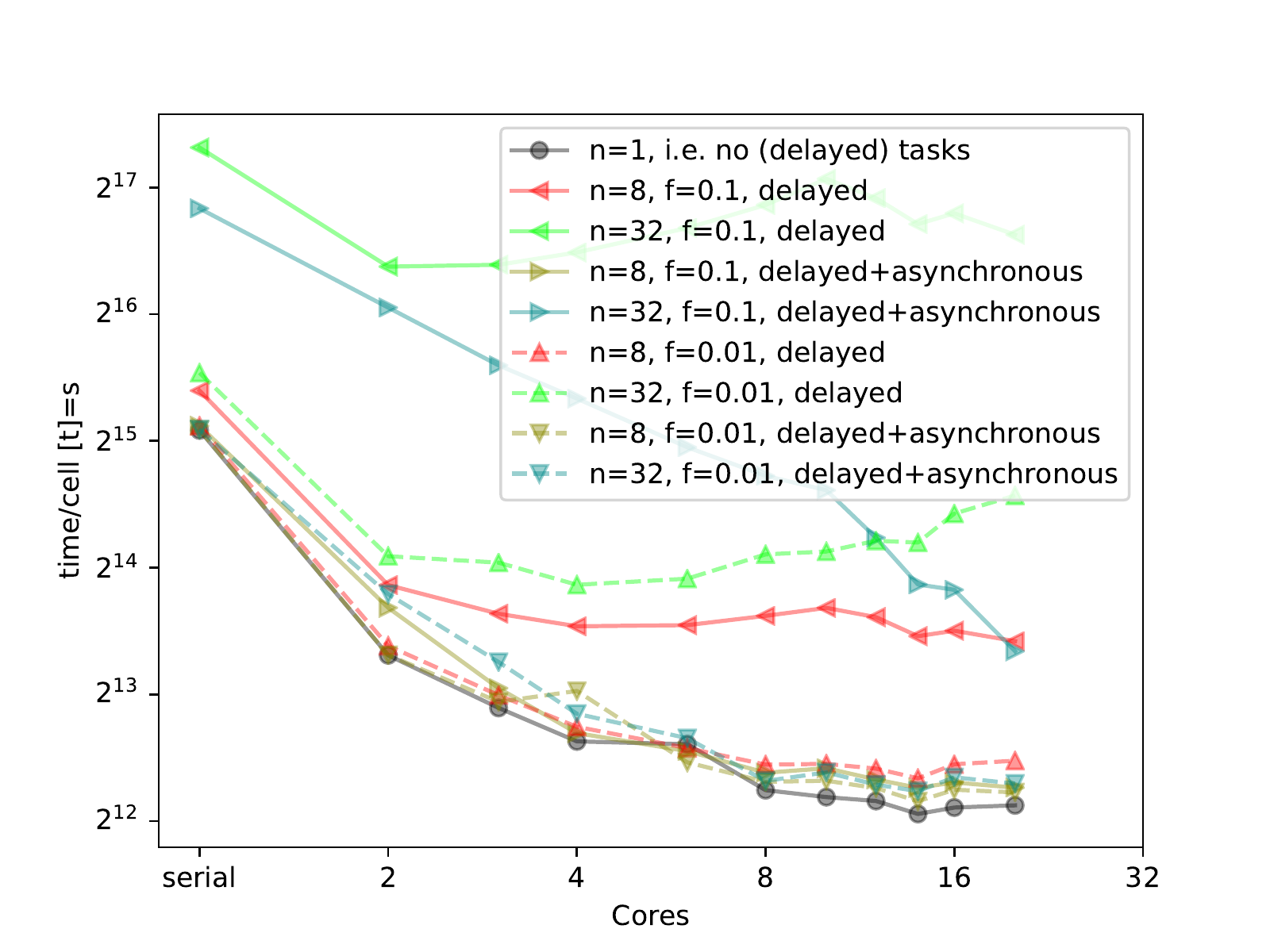}
 \end{center}
 \caption{
  Runtime per grid sweep for one discretisation with various
  integration/tasking configurations.
  Small grid with $h \leq 0.1 $ (left) vs.~slightly bigger grid with $h \leq
  0.005$ (right).
  \label{figure:scalability}
 }
\end{figure}

We wrap up our experiments with simple single node studies---the compression and
tasking paradigm has sole single node effect. 
The experiments run through a series of setups \deleted[id=R1]{always} per
tested grid.
First, we assess the pure scalability of the code without any delayed
integration and furthermore fix $n=1$.
Next, we prescribe $n>1$ and make the code yield $f \in \{0.1, 0.01\}$
integration tasks per cell, i.e.~between one and ten percent of the cells spawn
tasks.
As pointed out before, this fraction in real applications is not fixed.
We fix it manually here to assess the impact on scalability of our idea.
Finally, we run each \deleted{of the} experiment with a
\replaced[id=R1]{delayed}{``delayed''} integration twice:
In the baseline, the synchronisation is a \replaced[id=R1]{preamble to}{part of}
the cell evaluation.
In the alternative test, there is no synchronisation, i.e.~we spawn the integration
and do not wait for the result actively at any point.
We work totally asynchronously.
The setup is chosen such that the task spawn pattern mitigates the situation for
high $\theta $ values, i.e.~all spawned tasks correspond to cells close to the
coordinate system axes.

\begin{figure}
 \begin{center} 
  \includegraphics[width=0.48\textwidth]{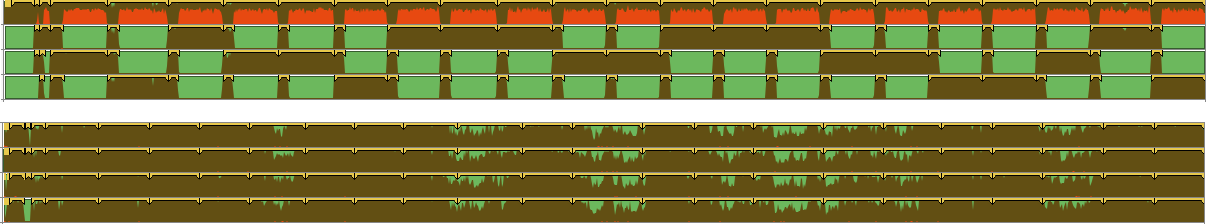}
 \end{center}
 \caption{
  Task distribution/placement for one setup with four cores. 
  Top: No delayed tasking is used but each cell immediately determines an
  improved operator before it continues.
  Bottom: We use an anarchic, i.e.~an asynchronous delayed operator integration.
  Brown labels denote compute work, red \replaced[id=R1]{is}{it} spinning
  (active waits), \replaced[id=R1]{green}{white} denotes idling.
  The graph is a \added[id=R1]{zoomed-in} snapshot
  \replaced[id=R1]{
  extracted from the total execution which spans 709.5s (a priori
  integration) vs.~428.3s (asynchronous, delayed integration).
  }{
  , i.e.~the x-axes do not illustrate the whole time to
  solution.
  The top simulation requires 709.5s vs.~428.3s for the asynchronous scheme.
  }
  \label{figure:task-placements}
 }
\end{figure}


The partitioning with $n=1$ yields reasonable performance
(Fig.~\ref{figure:scalability}).
This obviously is a ``flawed'' setup from a mathematics point of view yet
assesses that the underlying solver in principle does scale.
As the workload is deterministic---it is hard-coded and does not use any
additional tasking---the setup also clarifies that any tasking with $n>1$ has to
yield an unbalanced workload.

With integration for a ratio $f$ of the cells, we indeed observe a deteriorated
scalability.
This is due to the fact that the high workload cells cluster along strong
$\theta $ variations.
We use a geometric decomposition of the domain before we deploy the grid to the
cores, and this decomposition tries to avoid disconnected partitions.
As a consequence, one or few cores only are responsible for all the
high-workload cells (Fig.~\ref{figure:task-placements}).
With the anarchic tasking\deleted[id=R1]{ where we never wait for integration
outcomes but ``hope'' that they drop in eventually}, we \deleted[id=R1]{indeed}
see that the scalability curve flattens out again and that we gain performance.
This difference is \replaced{greater with}{the higher the} higher \deleted{the }workload per integration and \replaced{with}{the}
higher \deleted{the }core count.

\begin{observation}
 The asynchronous, delayed element integration helps to regain some scalability
 for unbalanced setups.
\end{observation}

\noindent
We observe that the cores that run out of work towards the end of their mesh
traversal pick up some of the pending integration tasks spawned by overbooked
colleague threads.
Heavy integration tasks automatically slot into ``idle'' time of the baseline
solver.
\added[id=R1]{
 The delayed, asynchronous integration yields a solver with a
 performance and scalability profile that is comparable to purely geometric
 multigrid where all operators are computed geometrically with $n=1$ sampling points
 per cell.
 Our scalability tests fix the fraction $f$ of cells that require an improved
 integration as well as $n$.
 They thus study only the scalability behaviour of one particular
 multigrid cycle.
 If we study the whole time-to-solution of a solver, we find that this
 behaviour typically translates into a walltime of around 2/3 of the baseline.
 Baseline here is an implementation that uses the exactly same code base yet
 realises the lazy evaluation pattern, i.e.~computes all operators prior to the
 first usage accurately.
 Walltime always comprises both assembly phase and solve phase.
}

\section{Conclusion and outlook}
\label{section:conclusion}

%
%
Matrix assembly becomes a non-negligible part of the 
overall cost of a multigrid solve
within many application landscapes.
It is thus important to optimise this step, too. 
Our proposed strategy to achieve this is three-fold: 
First, we abandon the idea to make the assembly fast. 
We \replaced[id=R1]{instead}{actually} make it more expensive, as we switch from
an a priori integration of the underlying weak form to an iterative approach.
\replaced[id=R1]{In our naive implementation without any hierarchical
numerical integration, this leads}{We instead approximate the
``just-about-right'' integration accuracy required, but this might lead} to redundant, repeated
\deleted[id=R1]{integral} evaluations\added[id=R1]{ of sampling
points}.
\replaced[id=R1]{Nevertheless,}{In return,} we obtain an assembly that
\deleted[id=R1]{first} yields rough approximations quickly.
It reduces algorithmic latency.
The actual accurate integration is then delivered in the background of the
actual computation.
We hide this computational cost.
Second, we introduce an anarchic variant of this delayed integration.
We thus ignore previously existing synchronisation points and obtain very high scalability.
Finally, we propose a compressed accuracy storage format where the data
footprint evolution follows the integration accuracy used.
In particular around the start time\deleted[id=R1]{ of a solve}, we operate with
low memory footprints and, hence, low memory bandwidth demands.

%
%
\replaced[id=R1]{
 Our assembly ideas face two extreme cases:
 Setups where expensive, algebraic operator integration is not
 required---well-shaped domains with constant $\epsilon $ for example give us
 setups where pure geometric multigrid succeeds---or setups for which an
 accurate operator computation is essential, as material parameters jump dramatically.
 For the former case,
 delayed, asynchronous stencil integration might introduce too much
 overhead, as it has to integrate each stencil at least twice to come to the
 conclusion that a more accurate integration is not required.
 In the latter case,
 it might yield a non-robust implementation.
} 
{The present paper has not been able to uncover big surprises:}
Even though we use inaccurate numerical approximations initially, we obtain
correct solutions\replaced[id=R1]{ }{---}with a reduced time-to-solution despite
the increased computational workload \added[id=R1]{for our tests}.
\replaced[id=R1]{Stabililty is to be expected}{
This is an obvious result} given that we focus on elliptic, linear
problems\deleted[id=R1]{ here}.
If properly implemented, these equations always yield the right solution agnostic of the
initial guesses provided to them.
Even if our anarchic, delayed approach introduces slightly incorrect initial
iterates, \deleted[id=R1]{it is thus clear that} we remain stable.
\deleted[id=R1]{The devil however is in the details:}
Our data \added[id=R1]{however} suggest that we have to be very careful with
dynamic termination criteria, and that it is very reasonable to anticipate if operator parts are
completely off.
Trying to correct iterates stemming from inaccurate discretisations
overly aggressively can cause \deleted[id=R1]{temporary} instabilities in
solution behaviour\added[id=R1]{---though only temporarily}.
We hence \added[id=R1]{propose to} either fall back to geometric features within
the mesh---this stabilises the convergence behaviour---or we skip multigrid updates rather than
to apply \replaced{updates that introduce error}{totally wrong updates}.

%
%
Through the lens of additive multigrid, we have
looked at worst case scenarios.
Inaccurate operators here should, in the theory, propagate immediately through
the whole multiscale system, and there are no inherent low-concurrency phases
where tasks naturally can slot in.
\added[id=R1]{
Even a dominance of cells that do not require iterative,
accurate integration has not penalised our runtime and footprint significantly.
The latter results from our low precision storage.
For the runtime, a single-accuracy
vs.~two-sample-point-accuracy integration does not make a massive runtime
difference if they are all per-cell, blocked (likely cache-local) operations,
and we can hence hide any compute overhead.
}
The fact \added[id=R1]{that} our ideas
have shown promise \deleted[id=R1]{here} suggests that they will also be
successful for real-world challenges \added[id=R1]{and other solver types}.
It is a natural next step to investigate into such more complex
setups---time stepping codes where multigrid is only a building block,
non-linear PDEs, or convection-dominated systems, for example---and study the impact
of our proposed techniques in more detail \added[id=R1]{for multiplicative 
solvers and more effective smoothers which might be more sensitive to inaccurate
stencils}.
One of the most appealing strategies in an era of reducing or stagnating
memory per core is our idea to store operator entries with truncated precision.
So far, we use this reduced precision only to store data.
On the long term, it is \replaced[id=R1]{natural}{a natural strategy} to exploit
this also for mixed or reduced precision computing.
This is \replaced[id=R1]{timely}{promising} as we are currently witnessing the
introduction of \deleted[id=R1]{more} reduced-precision compute formats due to the success of
machine learning applications.

\ifthenelse{\boolean{arxiv}}{
  \bibliographystyle{siam}
}{}

\bibliography{paper}


\end{document}